\pdfoutput=1
\documentclass[aps,amsmath,preprint,superscriptaddress,nofootinbib]{revtex4-1}  
\usepackage{graphicx}
\usepackage{color}
\usepackage{amsmath,amssymb}	
\usepackage{hyperref}
\usepackage{setspace}
\usepackage{comment}
\usepackage{array}  
\usepackage{cancel}
\usepackage{enumitem}
\usepackage{mathrsfs}
\usepackage{physics}
\usepackage{todonotes}
\usepackage{subcaption}
\usepackage{ulem}

\begin{document}
\newcolumntype{Y}{>{\centering\arraybackslash}p{23pt}} 


\preprint{IPMU19-0157}

\title{New Type of String Solutions with Long Range Forces}

\author{Takashi Hiramatsu}
\email[e-mail: ]{hiramatz@icrr.u-tokyo.ac.jp}
\affiliation{ICRR, The University of Tokyo, Kashiwa, Chiba 277-8582, Japan}
\author{Masahiro Ibe}
\email[e-mail: ]{ibe@icrr.u-tokyo.ac.jp}
\affiliation{ICRR, The University of Tokyo, Kashiwa, Chiba 277-8582, Japan}
\affiliation{Kavli IPMU (WPI), UTIAS, The University of Tokyo, Kashiwa, Chiba 277-8583, Japan}
\author{Motoo Suzuki}
\email[e-mail: ]{m0t@icrr.u-tokyo.ac.jp}
\affiliation{Tsung-Dao Lee Institute, Shanghai Jiao Tong University, Shanghai 200240, China}
\affiliation{ICRR, The University of Tokyo, Kashiwa, Chiba 277-8582, Japan}
\date{\today}
\begin{abstract}
We explore the formation and the evolution of the string network in the Abelian Higgs model with two complex scalar fields. 
A special feature of the model is that it
possesses a global $U(1)$
symmetry in addition to the $U(1)$ gauge symmetry.
Both symmetries are spontaneously broken by the vacuum 
expectation values of the two complex scalar fields. 
As we will show the dynamics 
of the string network is quite rich 
compared with that in the ordinary Abelian Higgs model with a single complex scalar field. 
In particular, we find a new type of string solutions in addition to the 
conventional Abrikosov-Nielsen-Olesen (local) string solution.
We call this the uncompensated string.
An isolated uncompensated string has a logarithmic divergent string tension as in the case of the global strings, although it is accompanied by a non-trivial gauge field configuration.
We also perform classical lattice simulations in the $2+1$ dimensional spacetime, which confirms
the formation of the uncompensated strings at the phase transition. 
We also find that most of the uncompensated strings evolve into the local strings at later time when the gauge charge of the scalar field with a smaller vacuum
expectation value is larger than that of the scalar field with a larger vacuum expectation value.

\end{abstract}

\maketitle


\newpage
\section{Introduction}
\label{sec:introduction}

Cosmic strings (also known as vortex solutions)
are one-dimensional topological defects 
which appear in various contexts of particle physics 
and condensed matter physics.
The simplest theoretical framework to describe 
the string (vortex) formation is the Abelian Higgs model,
where the $U(1)$ gauge symmetry, $U(1)_{local}$, is spontaneously broken 
by a vacuum expectation value (VEV) of a complex scalar field. 
The string appearing in this model is called the Abrikosov-Nielsen-Olesen
(ANO) string~\cite{Nielsen:1973cs}.
At the phase transition of the $U(1)_{local}$ breaking, 
the strings are formed 
and make up a web-like structure, so-called the string network 
(see $e.g.$ the textbook~\cite{Vilenkin:2000jqa}). 
Numerical simulations have widely investigated the properties of the interactions and the evolution of the string network (see $e.g.$~\cite{Albrecht:1984xv,Albrecht:1989mk, Bennett:1989yp, Allen:1990tv, Vincent:1996rb,Martins:2005es, Ringeval:2005kr, Olum:2006ix, Fraisse:2007nu, BlancoPillado:2011dq,Vincent:1997cx,Moore:2001px,Bettencourt:1994kf,Bettencourt:1996qe, Salmi:2007ah,Achucarro:2006es,Pen:1997ae,Durrer:1998rw,Yamaguchi:1998gx,Yamaguchi:1999dy,Yamaguchi:1999yp,Yamaguchi:2002zv,Yamaguchi:2002sh,Cui:2007js,Hiramatsu:2013tga,Hindmarsh:2018wkp}).

In this paper, we discuss the formation and the evolution of the string network 
in the Abelian Higgs model with two complex scalar fields.
In particular, we consider a model with an additional global $U(1)$ symmetry, 
$U(1)_{global}$.
Such an additional global symmetry naturally arises when the 
ratio of the absolute $U(1)_{local}$ charges of the two complex scalar fields is larger than $3$ (or less than $1/3$).
In this case, there is no gauge-invariant interaction term which breaks $U(1)_{global}$ 
at the renormalizable level, and hence, it emerges as an accidental symmetry.
This class of models has been considered, for instance, in the context of the 
axion models where the Peccei-Quinn symmetry~\cite{Peccei:1977hh,Peccei:1977ur,Weinberg:1977ma,Wilczek:1977pj} is protected  by (abelian) gauge  symmetries~\cite{Lazarides:1982tw,Barr:1982uj,Choi:1985iv,Fukuda:2017ylt,Fukuda:2018oco,Ibe:2018hir} from the quantum gravity effects~\cite{Hawking:1987mz,Lavrelashvili:1987jg,Giddings:1988cx,Coleman:1988tj,Gilbert:1989nq,Banks:2010zn}.
This type of models is also discussed in condensed matter physics to understand the physics of high-temperature superconductivity~\cite{Lee:2006zzc}.%
\footnote{See also Refs.~\cite{Chernodub:2005qh,Chernodub:2005ts,Bock:2007vx,Babaev:2009,Catelani:2009,Forgacs:2016dby,Forgacs:2016iva}.}

As we will see, 
the dynamics of the string network 
is much richer compared with that in the ordinary Abelian Higgs model with a single complex scalar field. 
In particular, we observe the existence of a new type of string solutions in addition to the ANO (local) string.
We call the new type of the string solutions, the ``uncompensated'' strings. 
Around an uncompensated string, the covariant derivatives of the complex scalar fields are not 
canceled by the gauge field configuration at the large distance from the core of the string.
Accordingly, an isolated uncompensated string has a logarithmically divergent string tension
as in the case of the global strings.
In the early universe, those strings can be formed at the phase transition, 
where the divergence is cutoff by a typical distance between the strings of the order of the Hubble length. 
Our classical lattice simulations (in the $2+1$ spacetime dimensions) confirm the formation 
of the uncompensated strings at the phase transition.

As a characteristic feature of this system, the strings with winding numbers larger than $1$ appear ubiquitously. 
This feature should be contrasted with the string formation in the Abelian Higgs model with a single complex scalar field
in which it is quite rare to have a cosmic string with a large winding number in the cosmological context.
We also find that the uncompensated strings have long-range repulsive and attractive forces.
With the long-range forces, 
we find that the uncompensated strings tend to combine into the local ANO strings 
in the evolution of the string network when the gauge charge of the scalar field with a smaller VEV is larger than that of the scalar field with a larger VEV.

The organization of the paper is as follows. 
In Sec.~\ref{sec:stringsolution}, we discuss the static string solutions analytically.
In Sec.~\ref{sec:simulation}, we show the results of the classical lattice simulations
of the formation and the evolution of the string network in the $2+1$ dimensional spacetime.
In Sec.~\ref{sec:interaction}, we derive the long-range force between the uncompensated 
string and the global string in an analytical way.
The final section is devoted to our conclusions and discussions.

\section{String Solutions in Abelian Higgs Model}
\label{sec:stringsolution}

In this section, we discuss the string solutions in the Abelian Higgs model 
with two charged scalar fields.
We assume that both the scalar fields obtain non-vanishing VEVs.
A specific feature of the model is that it possesses
the $U(1)_{global}$ symmetry
in addition to 
the $U(1)_{local}$ symmetry.
We here discuss the static field configurations in an analytical way with the expansion of the universe being neglected.

\subsection{Model}

The action of the model is given by%
\footnote{This paper employs the metric $(-,+,+,+)$. }
%
\begin{align}
S=-\int d^4x\,(\mathcal{L}_\Phi+\mathcal{L}_A)\ ,
\end{align}
%
with the Lagrangian densities
%
\begin{align}
\mathcal{L}_\Phi&=(\mathscr{D}_\mu\phi_1)^*\mathscr{D}^\mu\phi_1
                + (\mathscr{D}_\mu\phi_2)^*\mathscr{D}^\mu\phi_2
                +V(\phi_1,\phi_2)\ ,\\
\mathcal{L}_A&=\frac{1}{4}F_{\mu\nu}F^{\mu\nu}.
\end{align}
%
Here, $\phi_n$ $(n=1,2)$ denote the two complex scalar fields and 
%
\begin{align}
F_{\mu\nu}=\partial_\mu A_\nu-\partial_\nu A_\mu\ ,
\end{align}
%
is the field strength of the gauge field $A_\mu$ of the $U(1)_{local}$ symmetry. 
The covariant derivative is defined by
%
\begin{align}
\mathscr{D}_\mu\phi_n:=\partial_\mu\phi_n-ieq_nA_\mu \phi_n\ ,
\end{align}
%
where $q_{n}$ is the charge of $\phi_n$, and $e$ is the gauge coupling constant.
In what follows, we normalize $q_{1}$ and $q_{2}$ so that they are relatively prime integers 
without loosing generality.

The scalar potential is taken to be
%
\begin{align}
\label{eq:potential}
V=\frac{\lambda_1}{4}(|\phi_1|^2-\eta_1^2)^2+\frac{\lambda_2}{4}(|\phi_2|^2-\eta_2^2)^2-\kappa(|\phi_1|^2-\eta_1^2)(|\phi_2|^2-\eta_2^2)\ ,
\end{align}
%
where $\lambda_{1,2}(>0)$ and $\kappa$ are real valued 
dimensionless coupling constants, and $\eta_{1,2}$ are real valued constants with a mass dimension.
This potential possesses two $U(1)$ symmetries 
under the phase rotations of the two complex scalar fields. 
One of the linear combination of these symmetries is 
identified as the $U(1)_{local}$ gauge symmetry,
while the other symmetry is a global $U(1)_{global}$ symmetry.
The global $U(1)_{global}$ symmetry naturally appears 
as an accidental symmetry in the renormalizable theory
when $|q_{1}| + |q_{2}|$ is larger than $4$.%
\footnote{For relatively prime integers $q_{1}$ and $q_{2}$,
the lowest dimensional $U(1)_{local}$ invariant but $U(1)_{global}$ breaking operators 
have the mass dimension of $|q_{1}| +|q_{2}|$. }

We consider the case with $\lambda_1\lambda_2>4\kappa^2$. In this case,
both the $U(1)$ symmetries are spontaneously broken by the VEVs~\cite{Saffin:2005cs},
%
\begin{align}
\label{eq:vacuum}
    \langle \phi_n \rangle = \eta_n \ , \quad (n = 1,2)\ .  
\end{align}
%
The Goldstone modes in this system are decomposed into 
the gauge-invariant Goldstone boson and the would-be Goldstone boson~\cite{Fukuda:2017ylt,Fukuda:2018oco}.
To see this, let us define the phase component fields $\tilde a_n$ 
$(n=1,2)$ by
%
\begin{align}
\label{eq:axialcomp}
\phi_1 = \frac{1}{\sqrt{2}}f_1\, e^{i \tilde a_1/f_1}\ , \quad 
\phi_2 =\frac{1}{\sqrt{2}}f_2\, e^{i \tilde a_2/f_2}\ ,
\end{align}
%
where the decay constants are given by $f_n = \sqrt{2} \eta_n$.
The domains of the phase component fields are given 
%
\begin{align}
\label{eq:domain}
\tilde{a}_{1}/f_1= [0, 2\pi)\ ,  \quad \tilde{a}_{2}/f_2 = [0, 2\pi)\ ,  
\end{align}
%
respectively.
The $U(1)_{local}$ gauge symmetry is realized 
by the shifts of $\tilde a_{n}$,
%
\begin{align}
\label{eq:shift}
\tilde{a}_{1}/f_1\to\tilde{a}_{1}/
f_1 + q_{1} \alpha \ ,
\quad \tilde{a}_{2}/f_2  \to\tilde{a}_{2}/f_2 + q_{2} \alpha \ ,
\end{align}
%
where $\alpha$ is a local parameter of the $U(1)_{local}$ transformation.

\begin{figure}[tbp]
\begin{center}
  \begin{minipage}{.5\linewidth}
  \includegraphics[width=\linewidth]{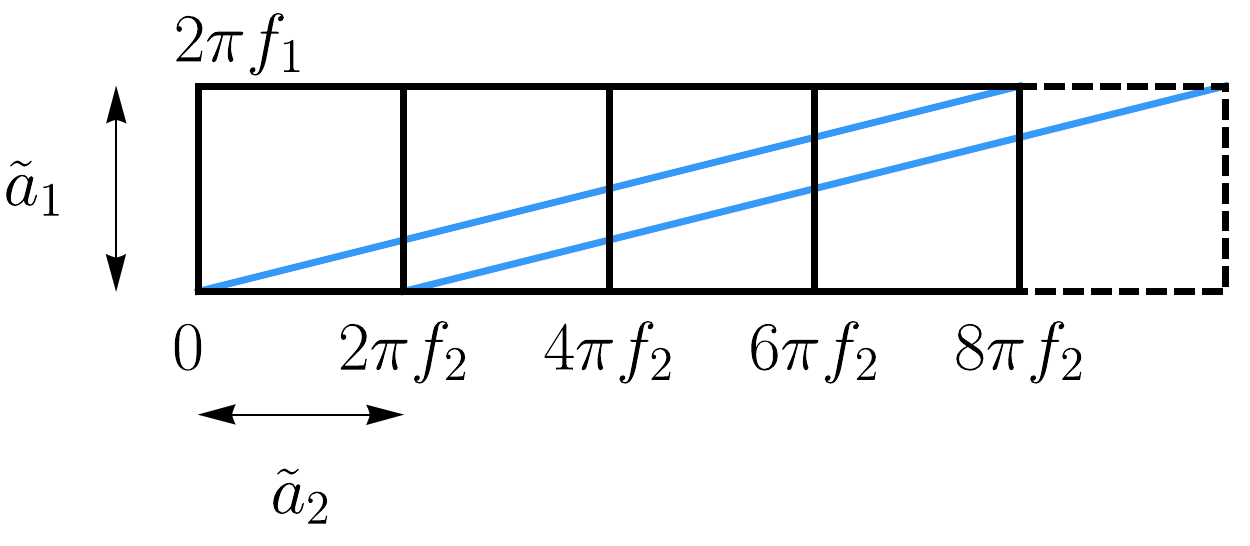}
 \end{minipage}
 \end{center}
  \caption{
\sl \small \raggedright 
A gauge orbit for the $U(1)_{local}$ gauge symmetry
for $(q_{1},q_{2}) = (1,4)$ (blue lines).
The orbit corresponds to the direction of the would-be Goldstone boson.
The direction perpendicular to the blue lines 
corresponds to the gauge-invariant Goldstone boson.
The domain of the gauge-invariant Goldstone boson
is given by the interval between the blue lines.
}
\label{fig:domain}
\end{figure}

The kinetic terms of $\tilde {a}_{n}$ are given by
%
\begin{align}
{\cal L} &= |\mathscr{D}_\mu \phi_1|^2 + |\mathscr{D}_\mu \phi_2|^2\notag \\
&= \frac{1}{2}(\partial \tilde a_1)^2 
+  \frac{1}{2}(\partial \tilde  a_2)^2 
- e A_\mu (q_{1} f_1  \partial^\mu \tilde a_1 + q_{2} f_2  \partial^\mu\tilde a_2)\notag \\
&-\frac{1}{2} e^2 (q_{1}^2 f^{2}_1 + q_{2}^2f^{2}_2 ) A_\mu^2 \notag\\
&= \frac{1}{2}(\partial  a)^2 + \frac{1}{2}m_A^2 \left(A_\mu - \frac{1}{m_A}\partial_\mu  b \right)^2\ .
\end{align}
%
In the final expression, we use the Goldstone bosons defined by
%
\begin{align}
\left(
\begin{array}{cc}
 a   \\
 b
\end{array}
\label{eq:decomp}
\right)=
\frac{1}{\sqrt{q_{1}^2f^{2}_1 
+q_{2}^2f^{2}_2 }}\left(
\begin{array}{cc}
q_{2} f_2   &  -q_{1} f_1   \\
q_{1} f_1  &   q_{2} f_2
\end{array}
\right)
\left(
\begin{array}{cc}
 \tilde a_1   \\
\tilde a_2
\end{array}
\right)\ .
\end{align}
%
The field $ b$ is the would-be Goldstone boson which is absorbed by 
the $U(1)_{local}$ gauge boson 
in the unitary gauge.
The mass of the gauge boson is given by,
%
\begin{align}
    m_A^2 = e^2\left(q_{1}^2 f^{2}_1 +  q_{2}^2 f^{2}_2\right)
\end{align}
%
The field $a$ is the Nambu-Goldstone boson of the global $U(1)_{global}$ symmetry,
which is invariant under the $U(1)_{local}$ symmetry (see Eq.\,\eqref{eq:shift}).

In Fig.~\ref{fig:domain}, we show the gauge orbits of $U(1)_{local}$ for $(q_{1},~q_{2})=(1,~4)$ as the blue lines.  
As seen in the figure, the shift of $b$ from $0$ to  $2\pi$ corresponds to the shift of $a_1$ from $0$ to $2\pi\,q_{1}$ and the shift of $a_2$ from $0$ to $2\pi\,q_{2}$.
The direction perpendicular to the blue lines 
corresponds to the gauge-invariant Goldstone boson, $a$.
The domain of $a$
is given by the interval between the blue lines,
%
\begin{align}
    a = [0,2\pi F_a)\ , \quad F_a = \frac{f_1f_2}
    {\sqrt{q^{2}_1 f^{2}_1 + 
    q^{2}_2f^{2}_2 }}\ .
    \label{eq:domainA}
\end{align}
%

\subsection{Field Equations of the Static String Solutions}

The Euler-Lagrange equations of $\phi_n$ and the gauge field are given by, 
%
\begin{align}
&\mathscr{D}^\mu\mathscr{D}_\mu\phi_n=\frac{\partial V}{\partial\phi^{*}_n}\ ,\\
&\partial_\mu F^{\mu\nu}={ie}\sum_{n=1}^{2}q_n
\left[\phi^{*}_n
\mathscr{D}^\nu\phi_n-(\mathscr{D}^\nu\phi_n)^*\phi_n
\right]\ .
\end{align}
%
Hereafter, we take the temporal gauge, $A_0=0$,
which reduces the field equations to
%
\begin{align}
\label{eq:eulers}
&\ddot{\phi}_n-\delta^{ij}\mathscr{D}_i\mathscr{D}_j\phi_n=-\frac{\partial V}{\partial\phi^{*}_n}\ ,\\
\label{eq:eulerA}
&\ddot{A}_k-\delta^{ij}\partial_iF_{jk}={-ie}\sum_n^2q_n
\left[\phi^{*}_n
\mathscr{D}_k\phi_n-(\mathscr{D}_k\phi_n)^*\phi_n
\right],
\end{align}
%
and yields the constraint equation,
%
\begin{align}
&\delta^{ij}\partial_i\dot{A}_j = -ie\sum_{n=1}^2[\phi_n^*\dot{\phi}_n-\phi_n\dot{\phi}_n^*].\label{eq:const}
\end{align}
%

The dot denotes the time-derivative.%
\footnote{The spacetime coordinate is defined 
by $(t, x_1,x_2,x_3) = (x^0, x^{1},x^2,x^3)$.}

As an ansatz for the static string solutions, we assume 
%
\begin{align}
\label{eq:string1}
\phi_1(r,\theta)&=\eta_1e^{in_1\theta}h_1(r)\ ,\\
\label{eq:string2}
\phi_2(r,\theta)&=\eta_2e^{in_2\theta}h_2(r)\ ,\\
\label{eq:gauge}
A_\theta(r)&=\frac{1}{e}\xi(r)\ ,~A_r=A_z=0\ ,
\end{align}
%
in the cylindrical coordinate where 
$(r,~\theta,~z)$  is the radial distance, the azimuth angle, and the height, respectively. 
The integers $n_{1,2}$ denote the winding numbers of the strings consisting of $\phi_1$ and $\phi_2$, respectively.

Under the ansatz, the field equations in Eqs.\,\eqref{eq:eulers} and \eqref{eq:eulerA} are reduced to
%
\begin{align}
\label{eq:elstring1}
&h_1''(R)+\frac{h_1'(R)}{R}-{\beta}_1 h_1(R)^3+\left({\beta}_1-\gamma_2\left(1-h_2(R)^2\right)-\frac{n_1^2}{R^2}\left(1-\frac{q_{1}}{n_1}\xi(R)\right)^2\right)h_1(R) =0\ ,\\
\label{eq:elstring2}
&h_2''(R)+\frac{h_2'(R)}{R}-{\beta}_2 h_2(R)^3+\left({\beta}_2-\gamma_1
\left(1-h_1(R)^2\right)-\frac{n_2^2}{R^2}\left(1-\frac{q_{2}}{n_2}\xi(R)\right)^2\right)h_2(R)
=0\ ,\\
\label{eq:elgauge}
&\xi''(R)-\frac{\xi'(R)}{R}-2c_1\left(\xi(R)-\frac{n_1}{q_{1}}\right)h_1(R)^2-2c_2\left(\xi(R)-\frac{n_2}{q_{2}}\right)h_2(R)^2=0\ .
\end{align}
%
Here, we rescale the radial coordinate $r$ to $R$ by $R=r/r_0$ where
%
\begin{align}
r_0=\frac{1}{e\sqrt{q_{1}^{2}\eta^{2}_1 +q_{1}^{2}\eta^{2}_2}} \ .
\end{align}
%
The dash in the field equations denotes the derivative with respect to $R$.
We also define
%
\begin{align}
\label{eq:cx}
 \beta_n=\frac{\lambda_{n}\eta^{2}_n}{2e^2\left({q_{1}^2
\eta^{2}_1+ q_{2}^2
\eta^{2}_2}\right)}\ ,\quad 
\gamma_n=\frac{\kappa\, \eta^{2}_n}{e^2\left({q_{1}^2
\eta^{2}_1+ q_{2}^2
\eta^{2}_2}\right)}\ ,\quad
c_n =\frac{q_n^2\eta^{2}_n}{q_{1}^2
\eta^{2}_1+ q_{2}^2
\eta^{2}_2}\ ,
\end{align}
%
for $n=1,2$. 

\subsection{Static String Solutions}

To obtain the static string solutions, 
we consider the following boundary conditions,
%
\begin{align}
h_1(R)=0\ ,~ h_2(R)=0\ ,~\xi(R)=0\ ,
\label{eq:boundary1}
\end{align}
 for $R \to 0$, and
%
\begin{align}
h_1(R)=1\ ,~h_2(R)=1 \ ,
\label{eq:boundary2}
\end{align} 
%
for $R\to \infty$.
Those boundary conditions are the same in the case of the static string solution of 
the Abelian Higgs model with one complex scalar field.

In the region of $R\to \infty$, $h_1(r)$ and $h_2(r)$ get close to unity, 
 the field equation of $\xi$ in Eq.~\eqref{eq:elgauge} becomes
%
\begin{align}
\xi(R)''-\frac{\xi(R)'}{R}-2\xi(R)+\frac{2c_1 n_1}{ q_{1}}+\frac{2c_2 n_2}{ q_{2}}=0\ ,
\end{align}
%
for $R\to \infty$.
This field equation has an asymptotic solution,
%
\begin{align}
\label{eq:ainf}
\xi_\infty = \frac{c_1 n_1}{ q_{1}}+\frac{c_2 n_2}{ q_{2}}
= \frac{n_1 q_{1}\eta^{2}_1
+n_2 q_{2}\eta^{2}_2
}{q_{1}^2
\eta^{2}_1+ q_{2}^2
\eta^{2}_2}\  ,
\end{align}
%
and hence, the gauge field becomes
%
\begin{align}
\label{eq:asymptoticA}
A_\theta=\frac{1}{e}
\frac{n_1 q_{1}\eta^{2}_1
+n_2 q_{2}\eta^{2}_2
}{q_{1}^2
\eta^{2}_1+ q_{2}^2
\eta^{2}_2} \ ,
\end{align}
%
for $R\to \infty$.

The asymptotic behaviors of the covariant derivatives
of $\phi_n$ are then given by
%
\begin{align}
\label{eq:cov1}
    \mathscr{D}_\theta \phi_1 &\to i
    \left(
    n_1 - q_{1}\frac{n_1 q_{1}\eta^{2}_1
+n_2 q_{2}\eta^{2}_2
}{q_{1}^2
\eta^{2}_1+ q_{2}^2
\eta^{2}_2} \right)
\phi_1\ ,\\
 \mathscr{D}_\theta \phi_2 &\to i
    \left(
    n_2 - q_{2}\frac{n_1 q_{1}\eta^{2}_1
+n_2 q_{2}\eta^{2}_2
}{q_{1}^2
\eta^{2}_1+ q_{2}^2
\eta^{2}_2} \right)
\phi_2\ .
\label{eq:cov2}
\end{align}
%
As the covariant derivatives contribute to the string tension via
%
\begin{align}
    \mu^2 \sim 2\pi \int r dr \frac{1}{r^2} |\mathscr{D}_\theta \phi_n|^2\ ,
\end{align}
%
the string tension diverges at $r\to \infty$ 
unless both $\mathscr{D}_\theta \phi_1$
and $\mathscr{D}_\theta \phi_2$ vanish.
Eqs.\,\eqref{eq:cov1} and \eqref{eq:cov2}
show that this is possible only when the winding numbers satisfy,
%
\begin{align}
\label{eq:compensated}
    n_1 = N_w \times q_{1}\ , 
    \quad 
     n_2 = N_w \times q_{2}\ , \quad N_w  \in \mathbb{Z} \ .
\end{align}
%
The string solution which satisfies 
Eq.\,\eqref{eq:compensated} has a finite string tension.
The string solution which does not satisfy 
Eq.\,\eqref{eq:compensated}, on the other hand, has
a logarithmically divergent tension as in the case of the global string.
In the following, we call the former strings the compensated (local) strings, while the latter the uncompensated strings.
It should be emphasized that the 
uncompensated strings are new-type of the string solutions which are absent in the 
conventional Abelian Higgs model with a single complex scalar field.

\begin{figure}
\begin{minipage}{.32\textwidth}
\centering
\subcaption{$(n_1,n_2)=(1,4)$}
\includegraphics[width=\linewidth]{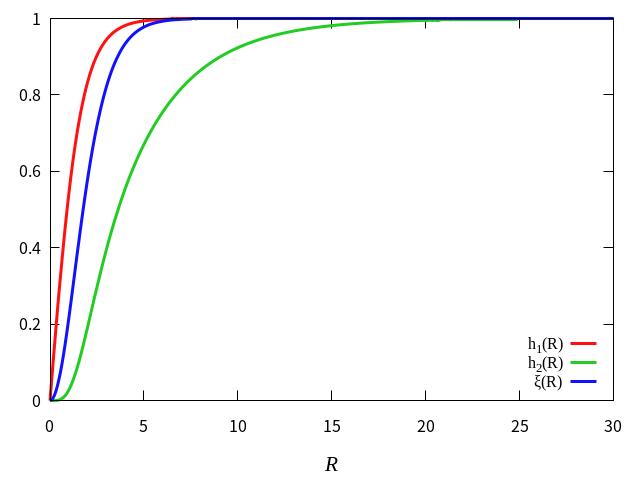} 
\end{minipage}
\begin{minipage}{.32\textwidth}
\centering
\subcaption{$(n_1,n_2)=(1,3)$}
\includegraphics[width=\linewidth]{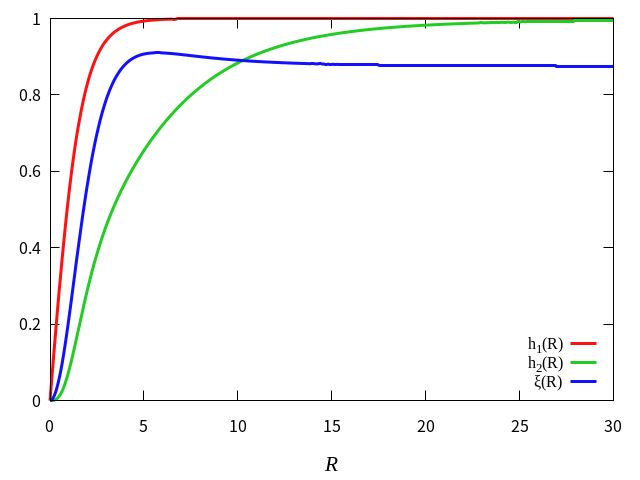}  
\end{minipage}
\begin{minipage}{.32\textwidth}
\centering
\subcaption{$(n_1,n_2)=(1,2)$}
\includegraphics[width=\linewidth]{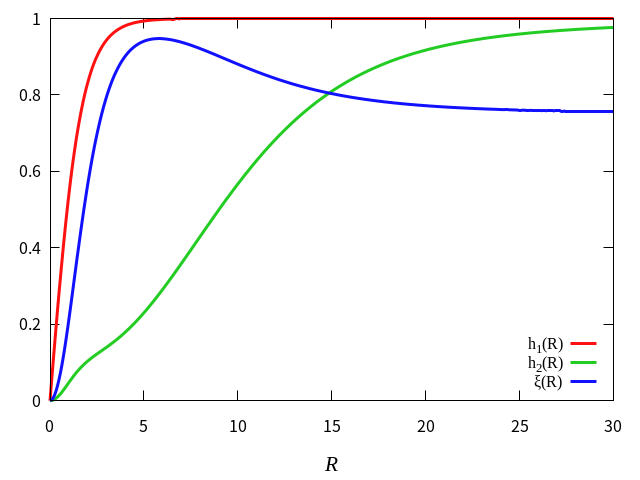}    
\end{minipage}\\
\vspace{2mm}
\begin{minipage}{.32\textwidth}
\centering
\subcaption{$(n_1,n_2)=(1,1)$}
\includegraphics[width=\linewidth]{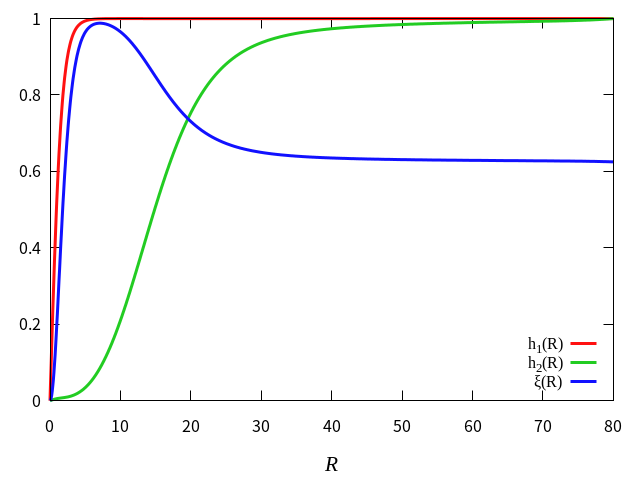}
\end{minipage}
\begin{minipage}{.32\textwidth}
\centering
\subcaption{$(n_1,n_2)=(1,0)$}
\includegraphics[width=\linewidth]{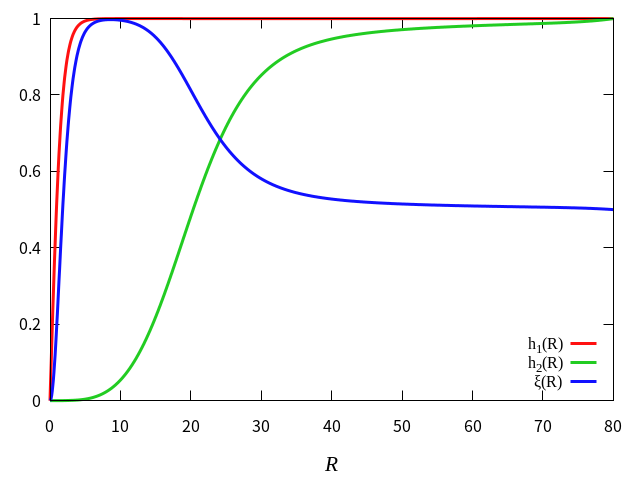}
\end{minipage}
\caption{
\sl \small\raggedright 
The static string solutions as 
a function of $R$ for given winding numbers.
In the figure, we take $q_{1}= 1$ and $q_{2}=4$
and $\eta_1/\eta_2 =4$.
The upper-left panel corresponds to the compensated string for $N_w = 1$.
The other strings are uncompensated strings.
Note that the figures in the first row are shown only up to $R\leq 30$
to emphases the structures around the string core,
while those in the second row is up to $R\leq 80$.
}
\label{fig:staticstrings}
\end{figure}

To show examples of the string profiles in our model, we solve the equations 
(\ref{eq:elstring1})-(\ref{eq:elgauge}) with 
the boundary conditions
given in Eqs.~(\ref{eq:boundary1})(\ref{eq:boundary2}) and (\ref{eq:ainf}).
We impose the asymptotic boundary conditions at $R=80$.
In Fig.~\ref{fig:staticstrings}, we show the static string solutions for various values of $(n_1,n_2)$ in the case of $(q_{1},q_{2})=(1,4)$. 
We take other parameters as {$e=1/\sqrt{2}$},  $\eta_1/\eta_2=4$, $\lambda_{1,2}= 1$,
and $\kappa = 0$.%
\footnote{For $\kappa\neq 0$, we have also numerically confirmed the existence of the static string solutions. One particular property is that the string core of $\phi_2$ becomes larger for larger $|\kappa|$ and the gauge field for the uncompensated strings takes $\xi=1$ over a wider interval in radius.}
In the figure, only the upper-left panel 
corresponds 
to the compensated string with $N_w = 1$.
The other configurations are the uncompensated strings.
The figure shows that the string core of $\phi_2$ becomes smaller for a larger $n_2$.
The figure also shows the gauge field configuration
converges to the asymptotic value in Eq.\,\eqref{eq:asymptoticA}.
We see that the gauge field configuration clings to the string solution of $\phi_1$ for a small $R$ 
and it converges to the asymptotic value at a large $R$.

The solution for  $n_2=0$ requires an explanation.
In this case, the condition, $h_2(r)=0$, at $R\to0$ is not required and we assume the Neumann boundary condition for $h_2$.
Even in this case, $h_2$ has a non-trivial configuration at around the string solution with $n_1 \neq 0$ (see Eq.\,\eqref{eq:elstring2}).

\begin{figure}[t!]
\begin{center}
  \begin{minipage}{.5\linewidth}
  \includegraphics[width=\linewidth]{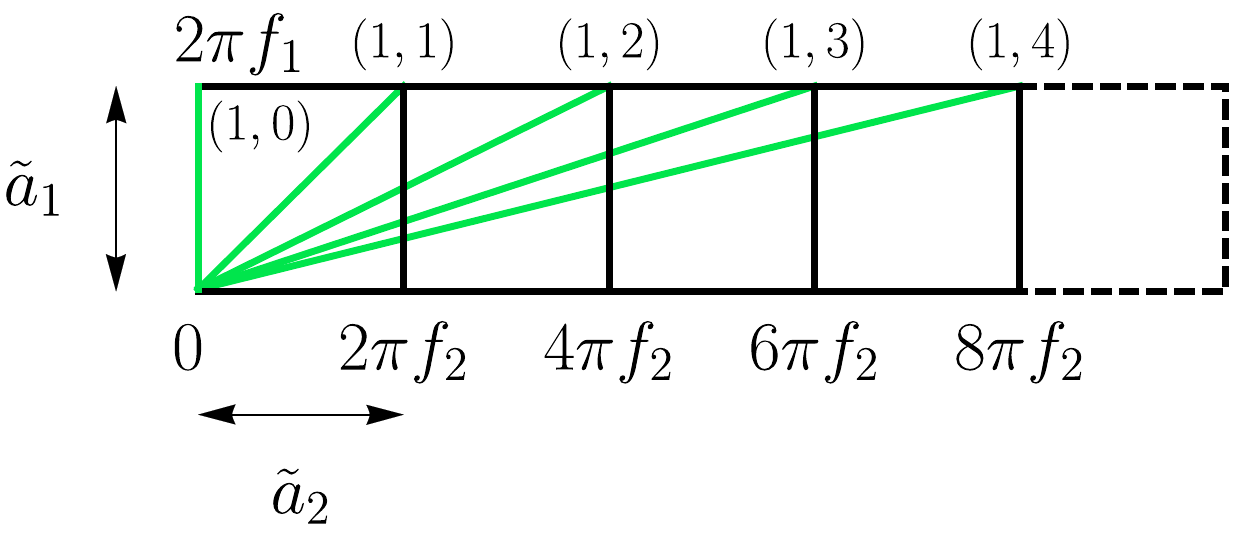}
 \end{minipage}
 \end{center}
  \caption{
\sl \small\raggedright 
The string solutions in the domain map of $(\tilde a_1, \tilde a_2)$ for 
given $(n_1,n_2)$.
The compensated string, $(1,4)$, corresponds 
to the direction of the orbit of the gauge transformation in Fig.~\ref{fig:domain}.
The uncompensated string winds not only
in the direction of the would-be Goldstone boson but also in the gauge-invariant Goldstone direction.
}
\label{fig:wind}
\end{figure}

In Fig.~\ref{fig:wind}, we show how each 
type of the string solutions winds 
in the domain map
of $(\tilde a_1, \tilde a_2)$
in the case of  $(q_{1}=1, q_{2}) =(1,4)$.
The figure shows that the configuration of the compensated 
string, i.e. $(n_1, n_2 )= (1,4)$, coincides
with the orbit of the gauge transformation 
(see Fig.~\ref{fig:domain}).
The uncompensated strings, on the other hand,
wind also in the direction of the gauge-invariant Goldstone boson.
In fact, the string configuration with $(n_1, n_2)$
winds the Goldstone direction as
\begin{align}
    \frac{a}{F_a} = \left(q_{2}n_1 - q_{1}n_2\right)\theta\ ,\quad  \theta \in [0,2\pi) \ ,
\end{align}
which vanishes only for the compensated strings (see Eq.\,\eqref{eq:decomp}).

\section{Classical Lattice Simulations}
\label{sec:simulation}

\subsection{Preparation for Simulations}
\label{sec:simulationsetup}

Here, let us summarize the 
setup of our numerical simulations.
To take into account the cosmic expansion
in the radiation dominated universe, 
we use the conformal time, $\tau$, with which 
the metric is given by,
%
\begin{align}
ds^2=a^2(\tau)(-d\tau^2+d\boldsymbol{x}^2)\ ,
\end{align}
%
where $a(\tau)$ is the scale factor.
In the expanding universe, the field equations are modified to
%
\begin{align}
\label{eq:eulers2}
&\ddot{\phi}_{n} + 2\mathcal{H}\dot{\phi}_{n} -\delta^{ij}\mathscr{D}_i\mathscr{D}_j\phi_n=-a^2V_{\phi^{*}_n}\ ,\\
\label{eq:eulerA2}
&\ddot{A}_k-\delta^{ij}\partial_iF_{jk}=-ia^2e\sum_n^2q_n
\left[\phi^{*}_n
\mathscr{D}_k\phi_n-(\mathscr{D}_k\phi_n)^*\phi_n
\right]\ ,
\end{align}
%
where $\mathcal{H}$ denotes the conformal Hubble parameter and the dot denotes the derivative with respect to the conformal time.

We also add the thermal mass term to the scalar potential
%
\begin{align}
    V_{\textrm{th}} 
    = \frac{T^2}{12}\left(
    \lambda_1 - \kappa + 3 e^2 q^{2}_1
    \right) |\phi_1|^2
    +
    \frac{T^2}{12}\left(
    \lambda_2 - \kappa + 3 e^2 q^{2}_2
    \right) |\phi_2|^2\ .
    \label{eq:thermalpot}
\end{align}
%
The thermal mass term stabilizes the symmetry 
enhancement point, $\phi_{1,2} = 0$, and hence,
the $U(1)_{local}\times U(1)_{global}$
symmetry is restored at the high temperature, $T \gg \eta_{1,2}$.

\begin{table}[t!]
\centering
\small{
\begin{tabular}{cc}
\hline
\hline
 \text{Grid size} &  $1024^2$\\
\text{Initial box size} &  $40H_{\rm in}^{-1}$ \\ 
\text{Final box size} &  $2H_{\rm fin}^{-1}$ \\
\text{Initial conformal time} & $2.01\eta^{-1}_1$ \\
\text{Final conformal time}& $40.2\eta^{-1}_1$\\
\text{Time step}& $1200$ \\ 
\hline
\hline
\end{tabular}
}
\caption{\raggedright Simulation parameters. $H_{\rm in}$ and $H_{\rm fin}$ are the Hubble parameter
at the initial time and the final time, respectively.}
\label{tab:uni}
\end{table}%

In the followings, we discuss the
formation/evolution of the strings for,
\begin{align}
\label{eq:parameters}
(q_{1},~q_{2})=(1,~4)\ ,~(\eta_1,~\eta_2)=(1,~0.25)\ ,~\kappa=0\ , \lambda_{1,2}= 1 \ , e = {\frac{1}{\sqrt{2}}}\ ,
\end{align}
as reference values. 
We also discuss the parameter dependencies
at the end of this section.
At the initial time, we impose the random values for $\phi_i(t_{\rm in},\boldsymbol{x})=\delta\phi(\boldsymbol{x})$ so that
the distributions of the scalar fluctuations are given as the Planck distribution with the temperature $T=\sqrt{3}\eta_1$.
Then we can determine $\dot{A}_i(t_{\rm in},\boldsymbol{x})$ by solving Eq.~(\ref{eq:const}) with the Fourier transformation. At this stage,
we can freely choose $A_i(t_{\rm in},\boldsymbol{x})$, so we simply assume $A_i(t_{\rm in},\boldsymbol{x})=0$.
Throughout the simulations, we assume the radiation-dominant universe.
We performed the simulations in a comoving box with periodic boundaries, and the time evolution is calculated by 
the Leap-Frog method and the spatial derivatives are approximated by the second-order finite difference.
The simulation parameters are summarized in Table~\ref{tab:uni}.%
\footnote{The spacial lattice spacing corresponds to $\sim 0.08\eta^{-1}_1$
at the initial time and $\sim 1.6\eta^{-1}_1$ at the final time.
}

\subsection{Formation of String Network}

\begin{figure}[!ht]
\begin{minipage}{.18\textwidth}
\centering
\subcaption*{$|\phi_1|$}
\includegraphics[width=1.2\textwidth]{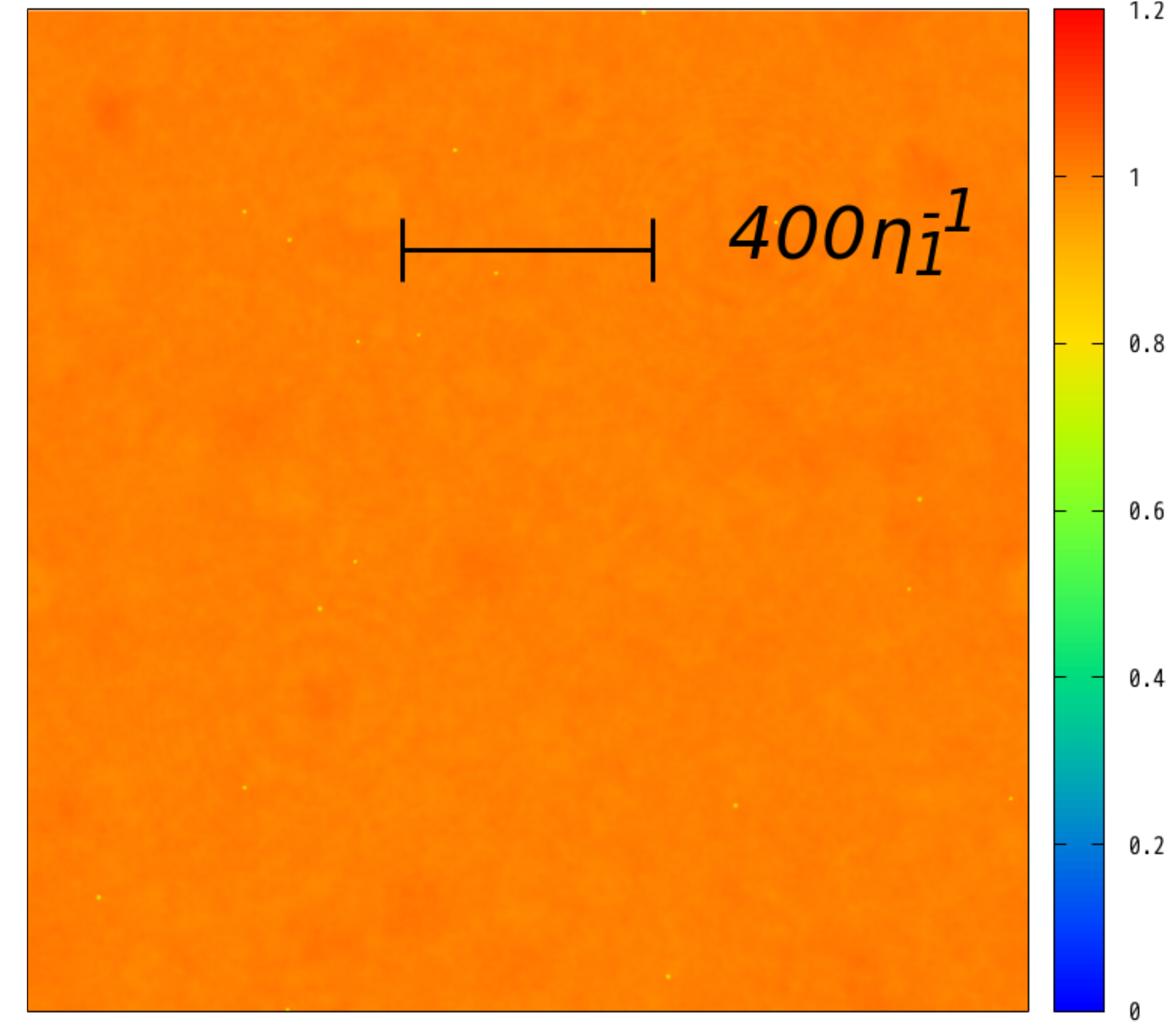} 
\end{minipage}
\begin{minipage}{.18\textwidth}
\centering
\subcaption*{$|\phi_2|$}
\includegraphics[width=1.2\linewidth]{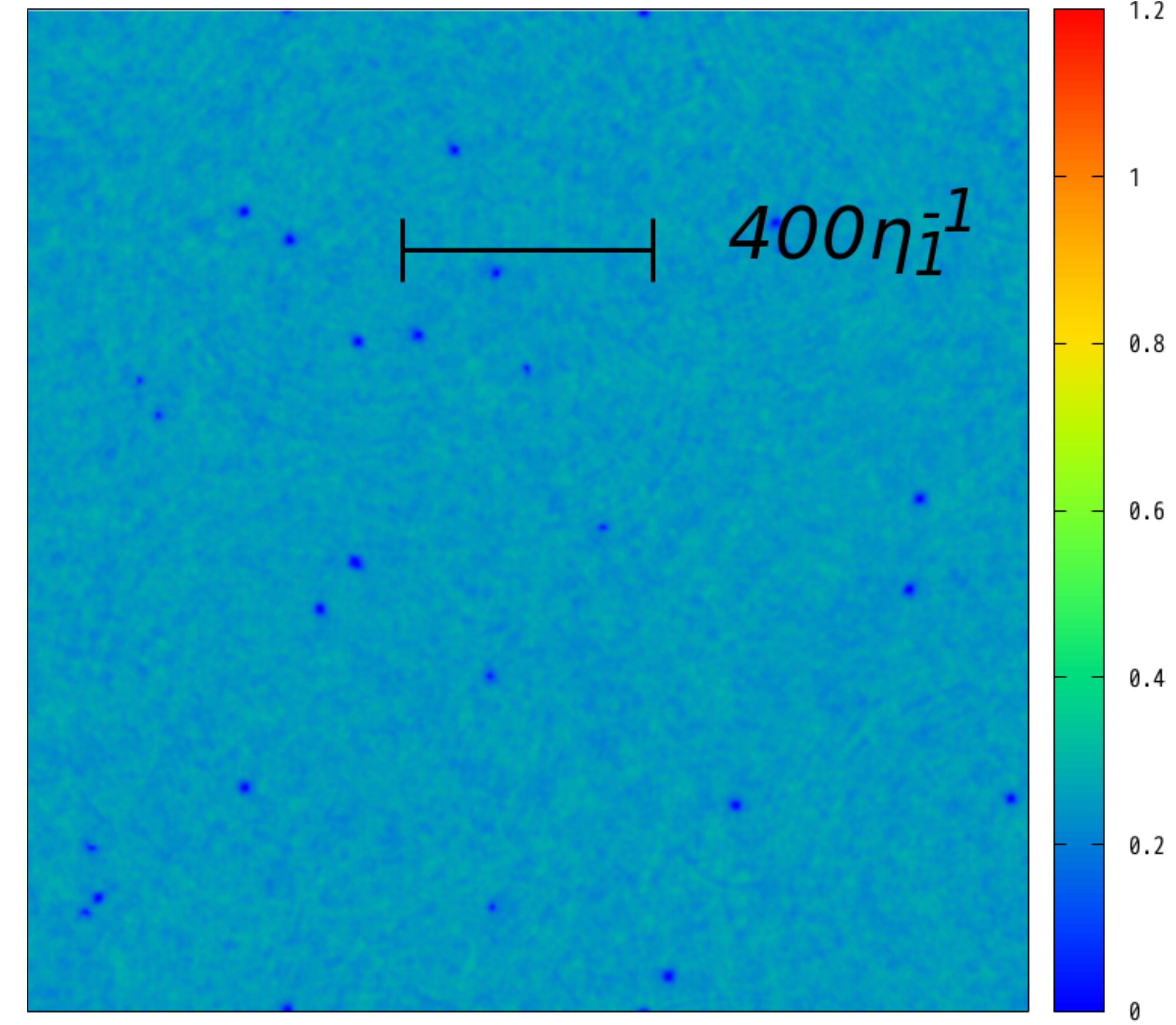}  
\end{minipage}
\hspace{2mm}
\begin{minipage}{.18\textwidth}
\centering
\subcaption*{$\arg \phi_1$}
\includegraphics[width=1.2\linewidth]{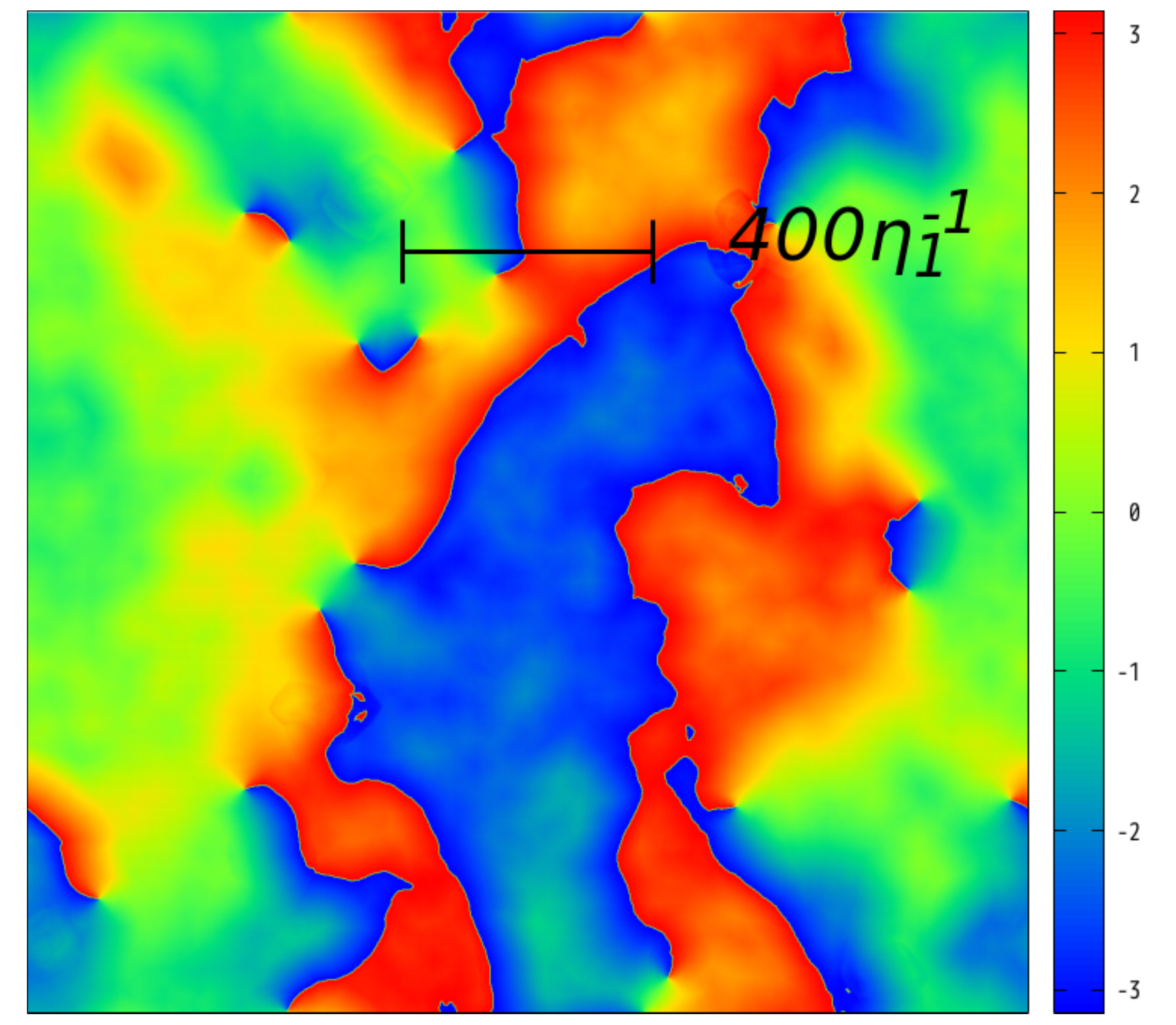}    
\end{minipage}
\begin{minipage}{.18\textwidth}
\centering
\subcaption*{$\arg \phi_2$}
\includegraphics[width=1.2\linewidth]{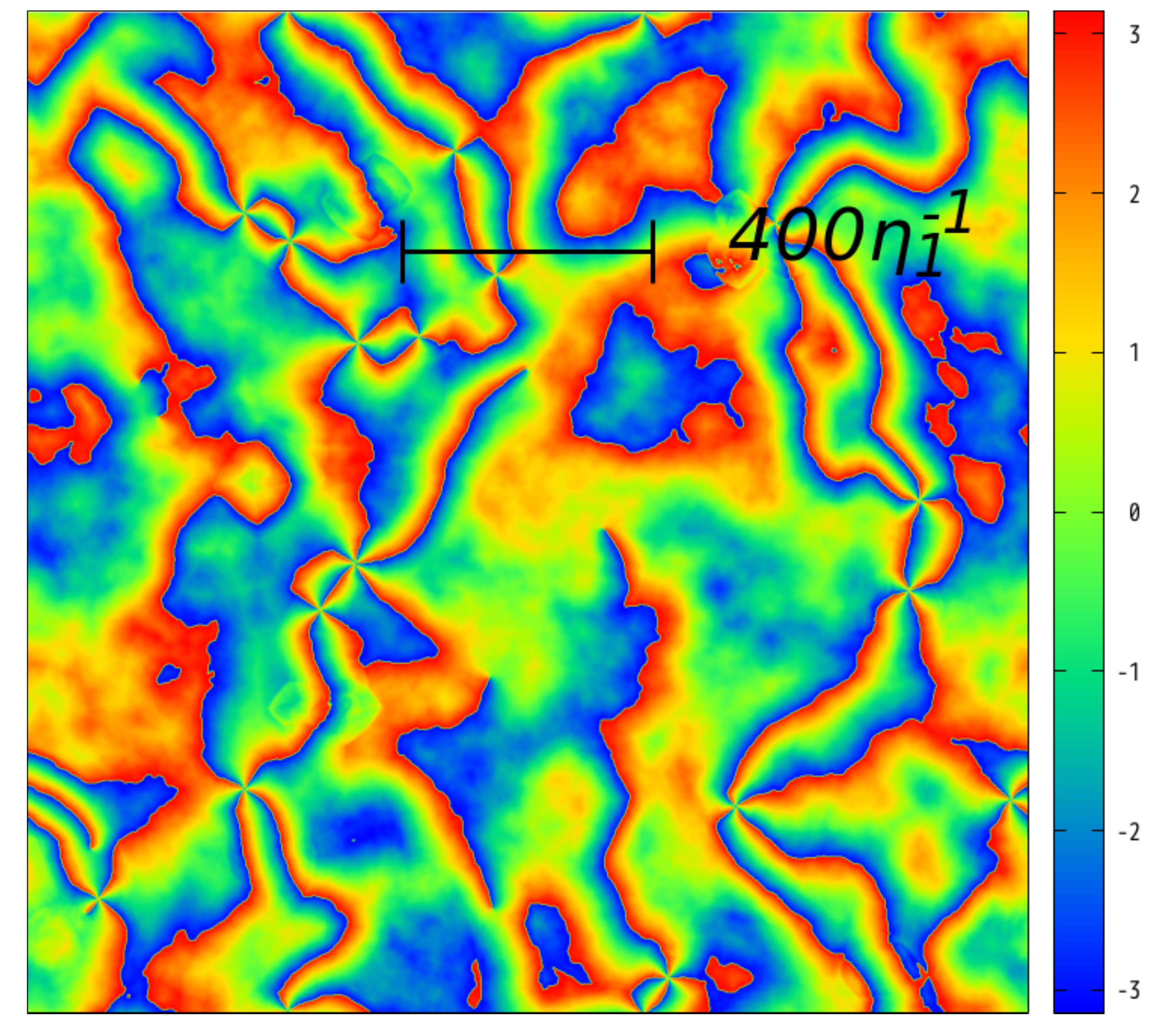}
\end{minipage}
\hspace{2mm}
\begin{minipage}{.18\textwidth}
\centering
\subcaption*{$(A_x,A_y)$}
\includegraphics[width=1.2\linewidth]{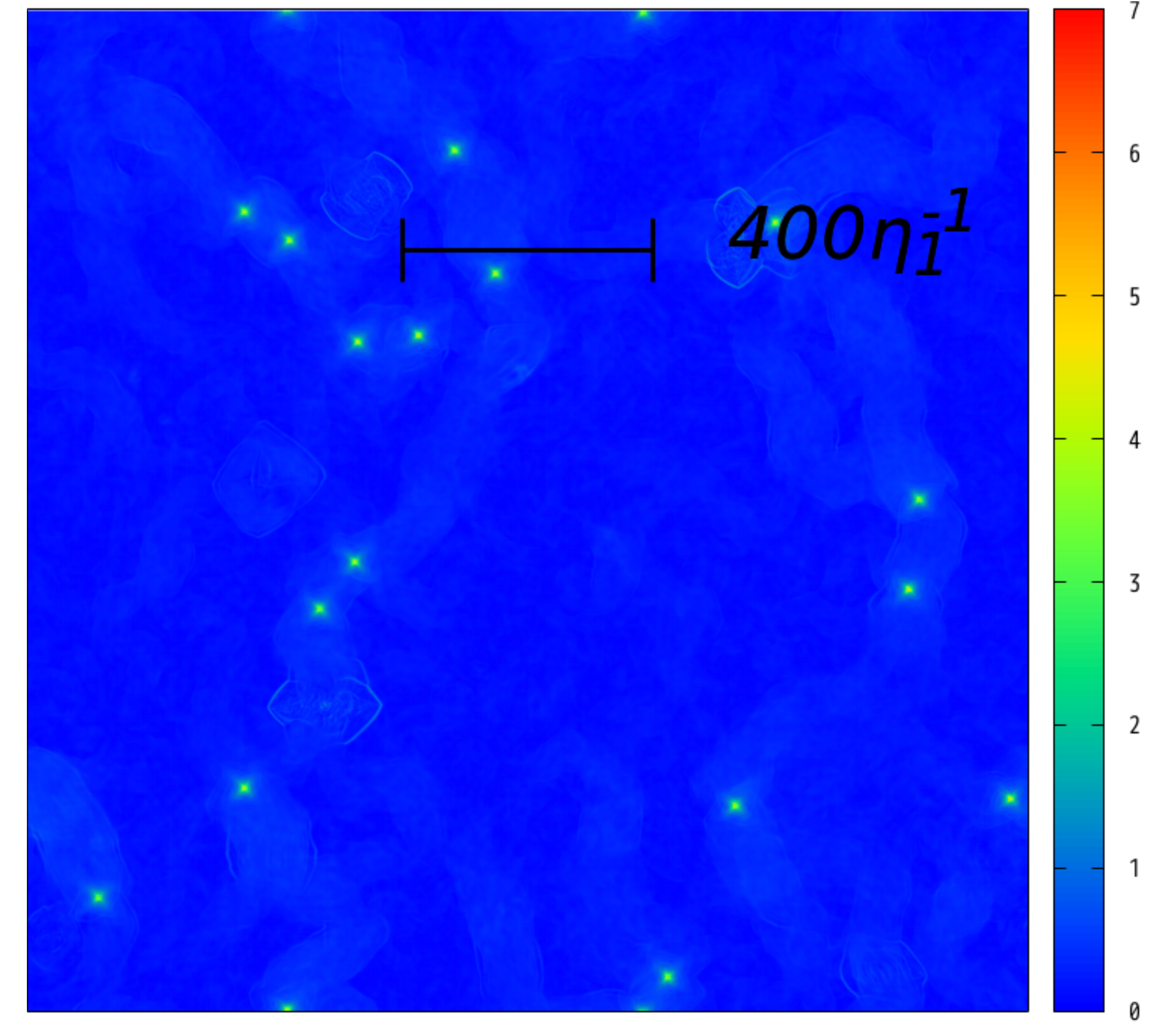}
\end{minipage}
\small{|| Spot 1 $(n_1,n_2) = (1,4)$ ||}\\
\begin{minipage}{.18\textwidth}
\centering
\includegraphics[width=1.2\textwidth]{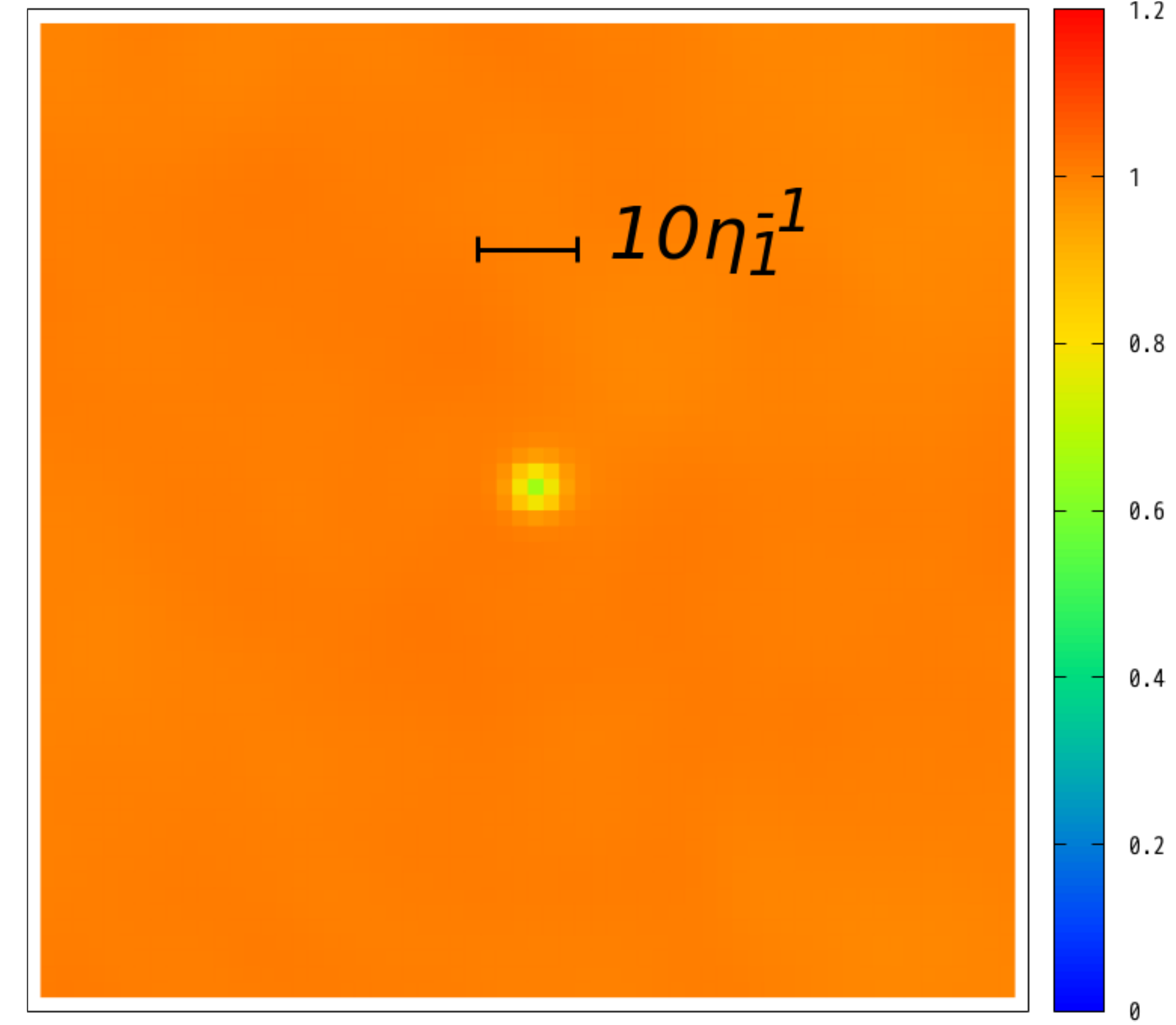} 
\end{minipage}
\begin{minipage}{.18\textwidth}
\centering
\includegraphics[width=1.2\linewidth]{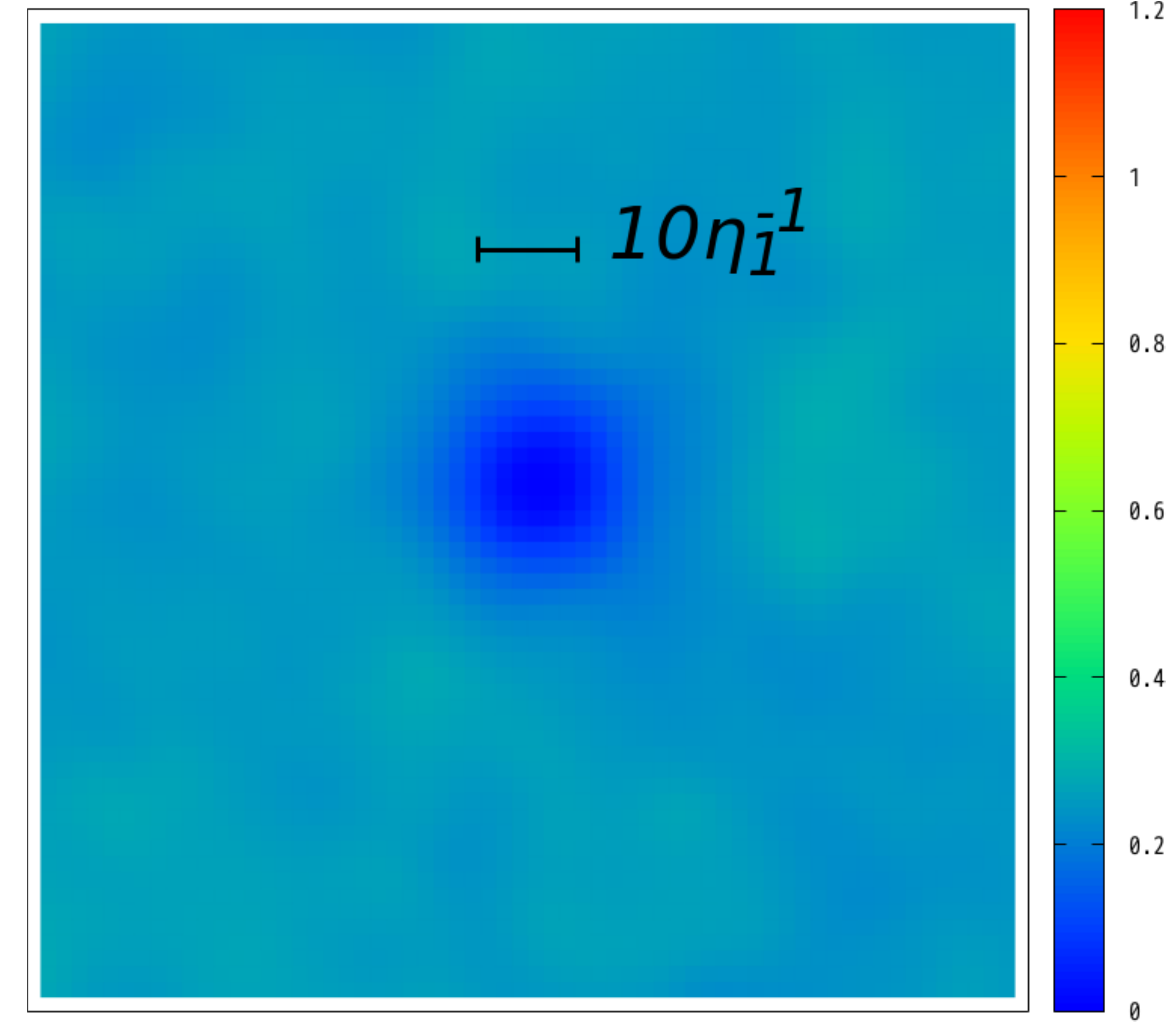}  
\end{minipage}
\hspace{2mm}
\begin{minipage}{.18\textwidth}
\centering
\includegraphics[width=1.2\linewidth]{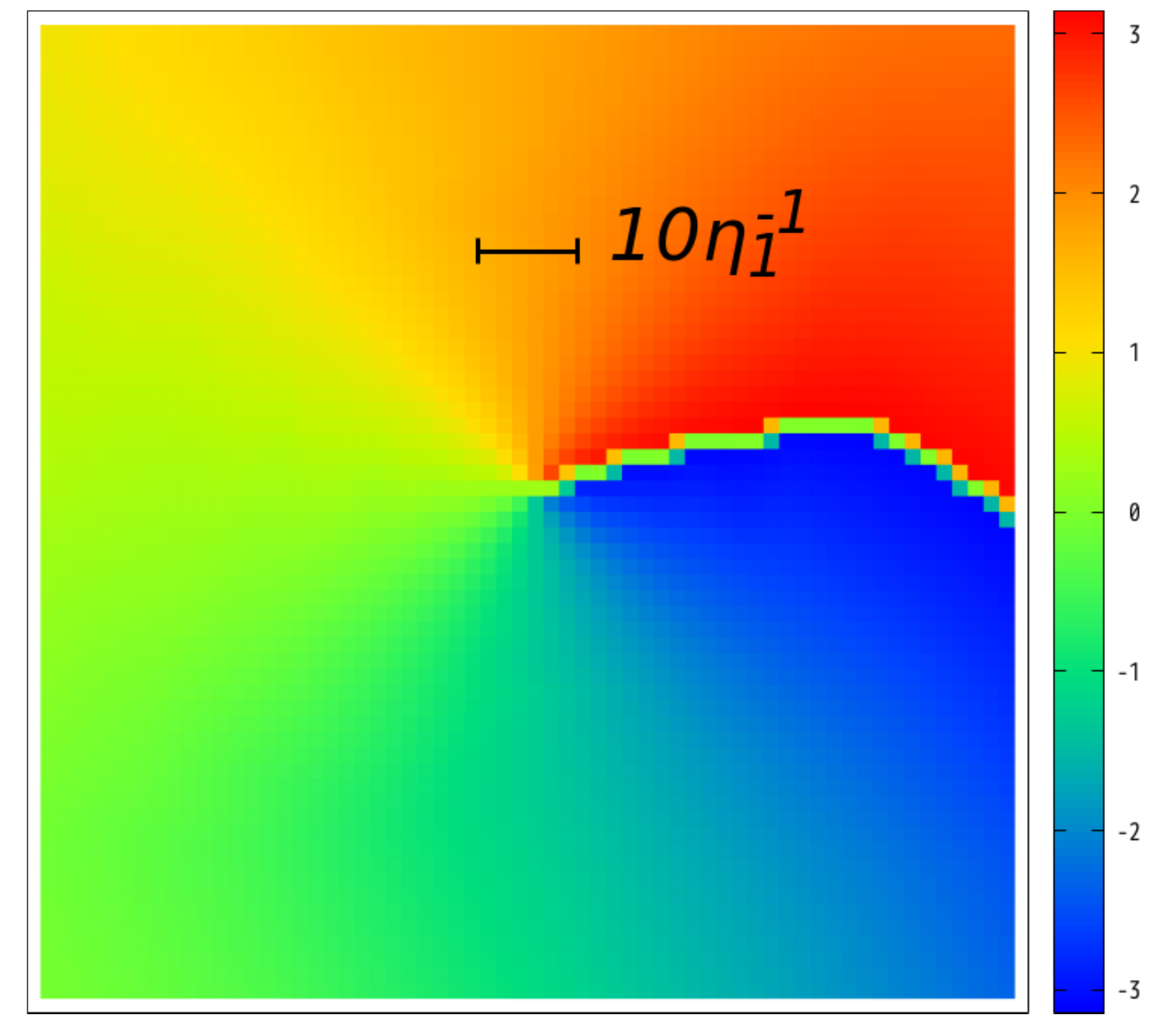}    
\end{minipage}
\begin{minipage}{.18\textwidth}
\centering
\includegraphics[width=1.2\linewidth]{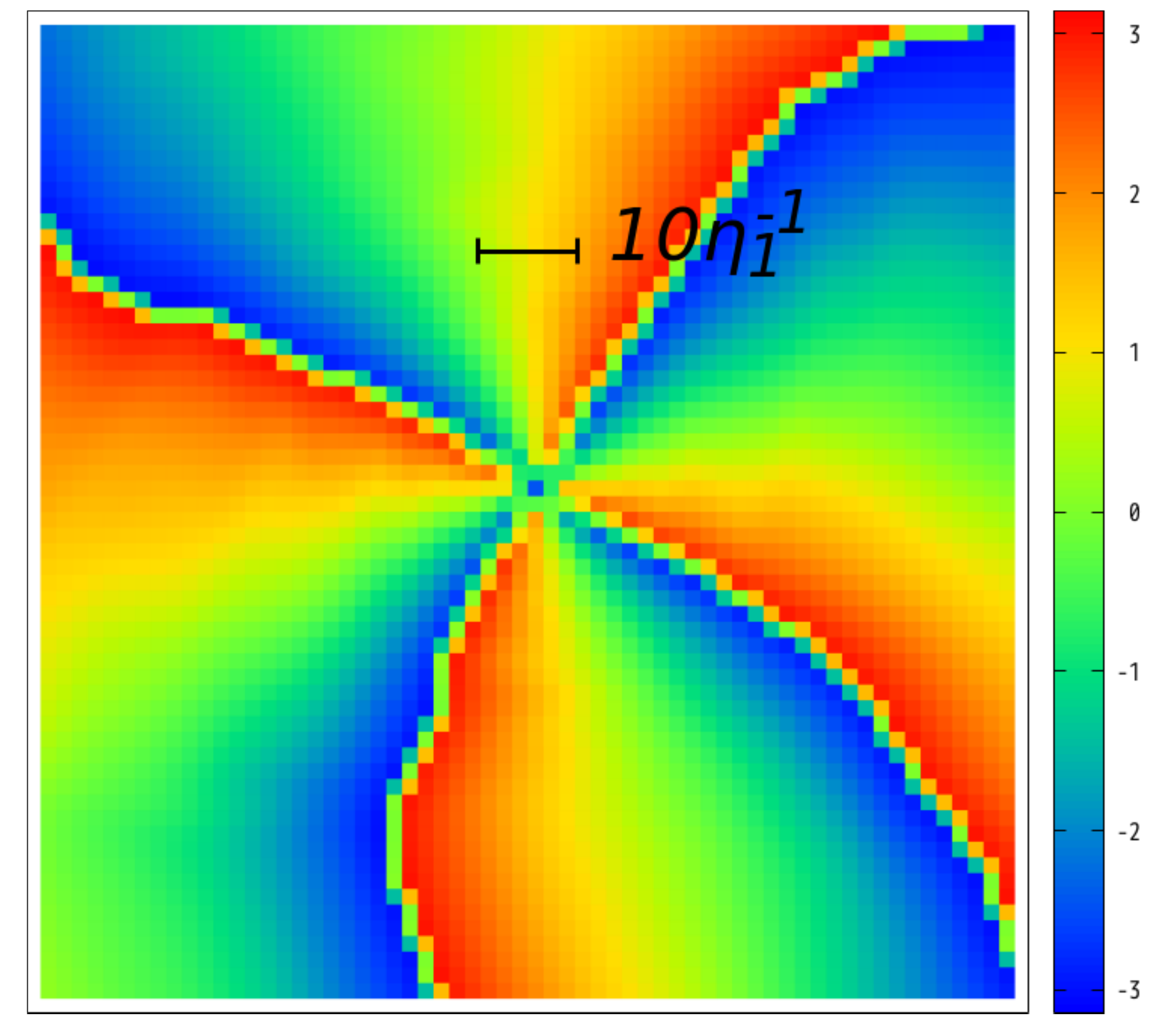}
\end{minipage}
\hspace{2mm}
\begin{minipage}{.175\textwidth}
\centering
\includegraphics[width=1.1\linewidth]{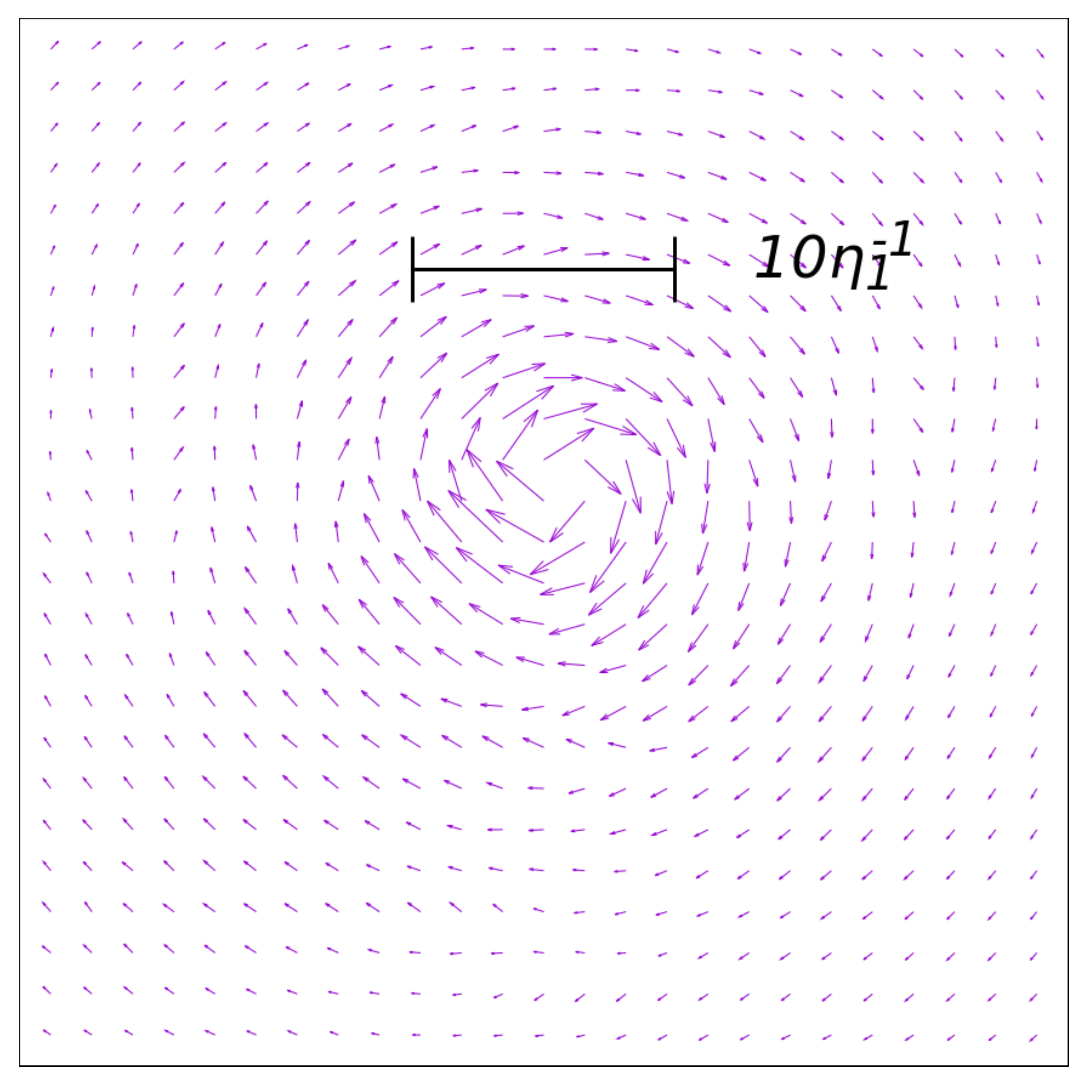}
\end{minipage}
\small{|| Spot 2 $(n_1,n_2) = (1,3)$ ||}\\
\begin{minipage}{.18\textwidth}
\centering
\includegraphics[width=1.2\textwidth]{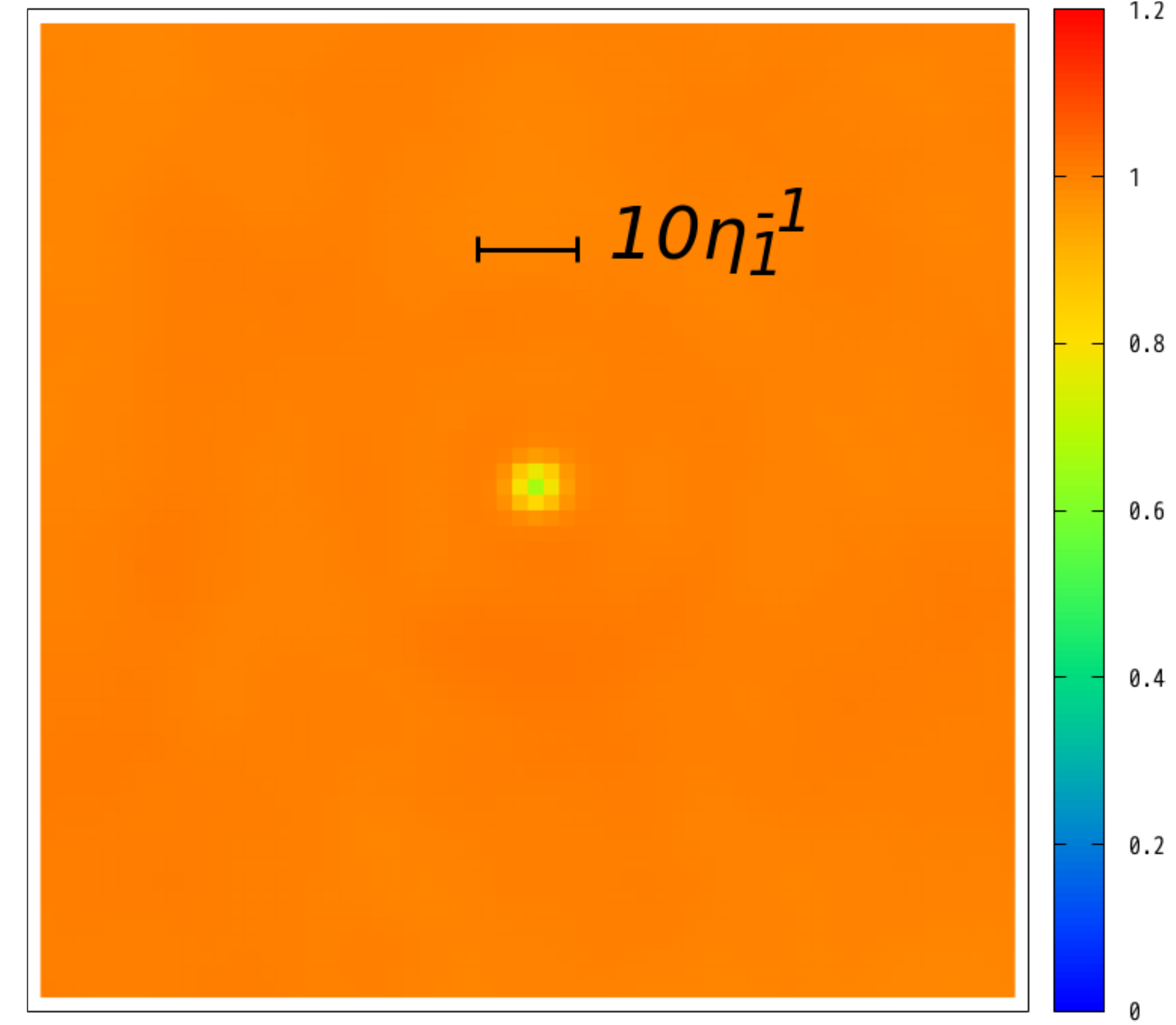} 
\end{minipage}
\begin{minipage}{.18\textwidth}
\centering
\includegraphics[width=1.2\linewidth]{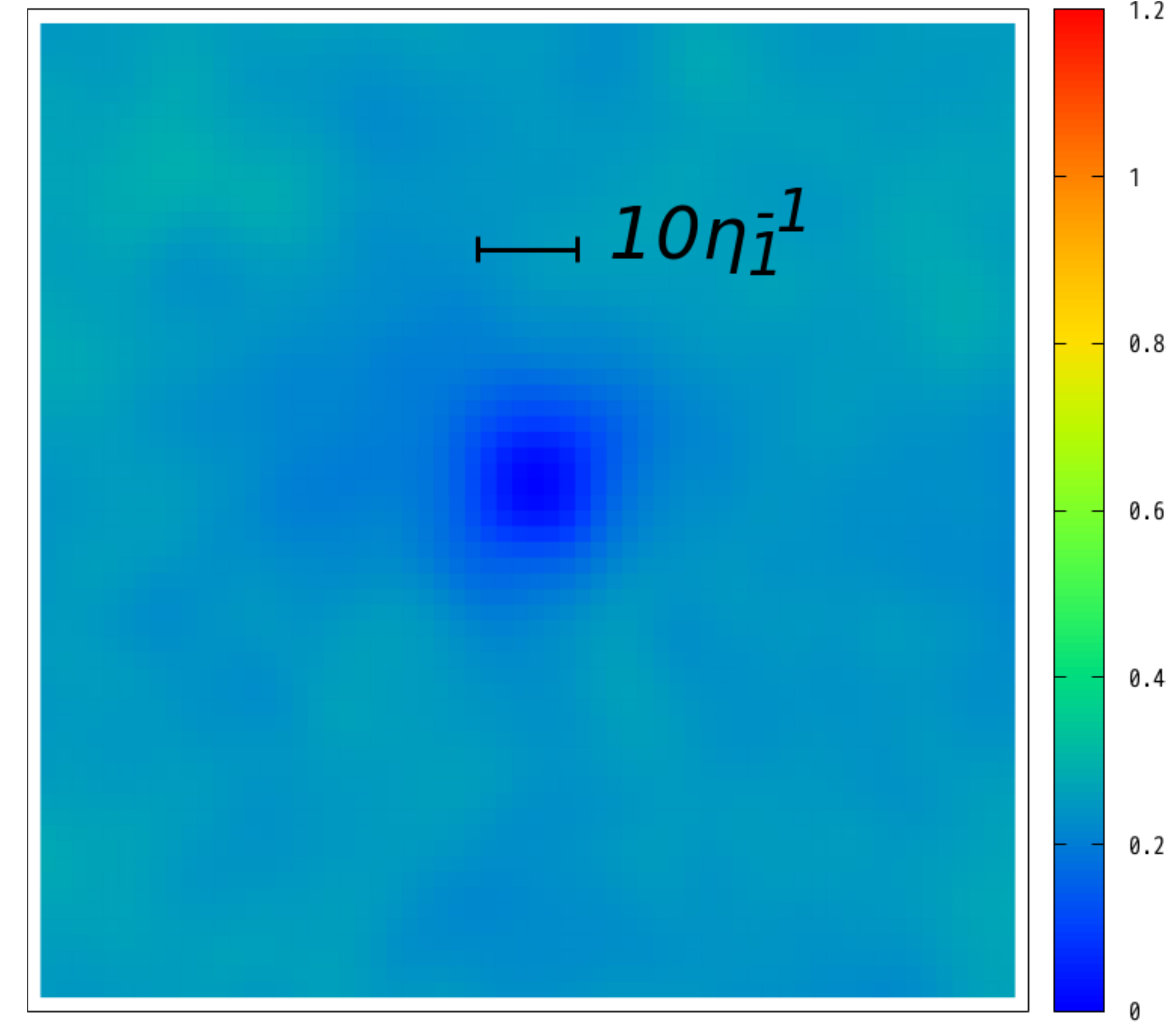}  
\end{minipage}
\hspace{2mm}
\begin{minipage}{.18\textwidth}
\centering
\includegraphics[width=1.2\linewidth]{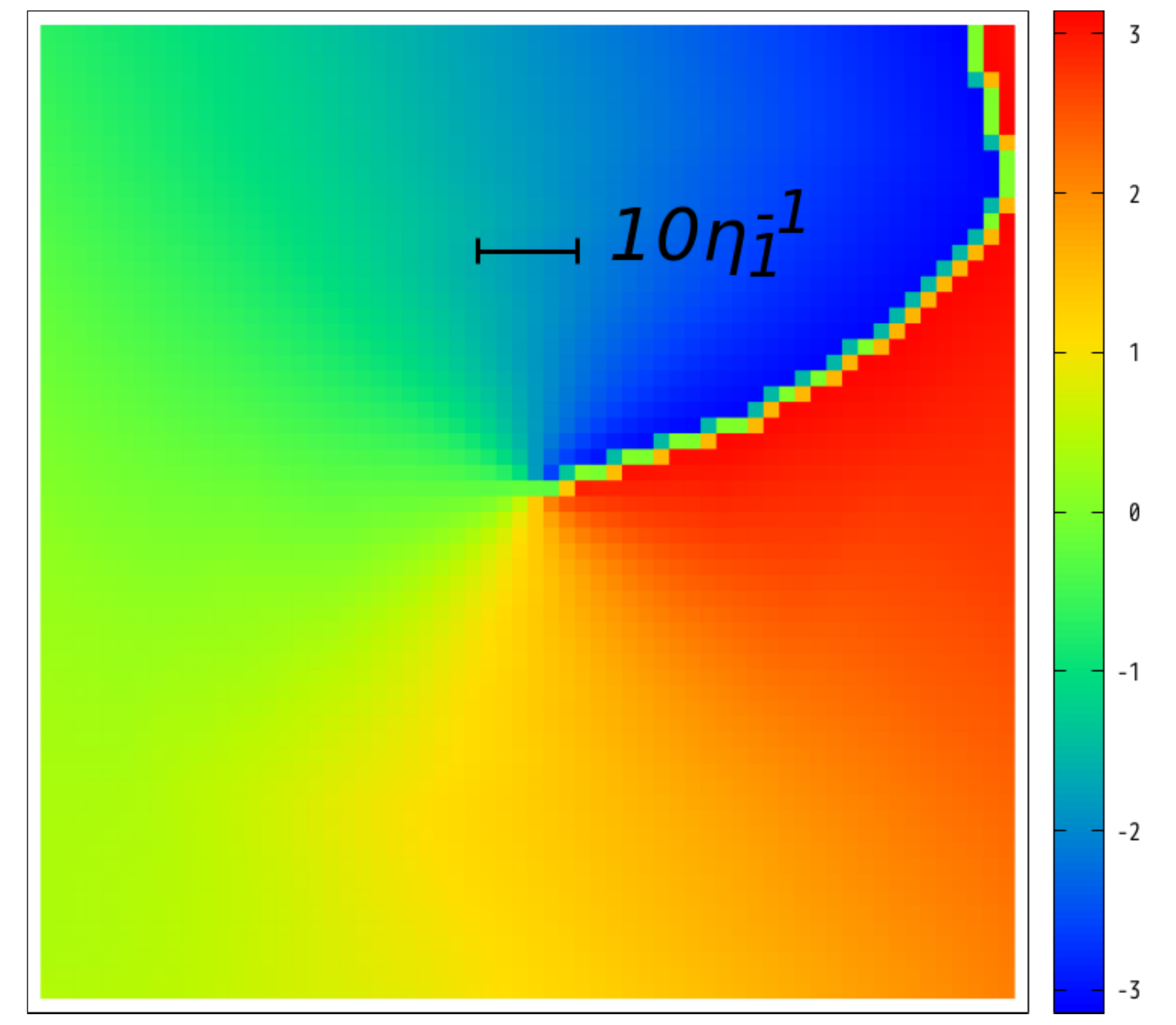}    
\end{minipage}
\begin{minipage}{.18\textwidth}
\centering
\includegraphics[width=1.2\linewidth]{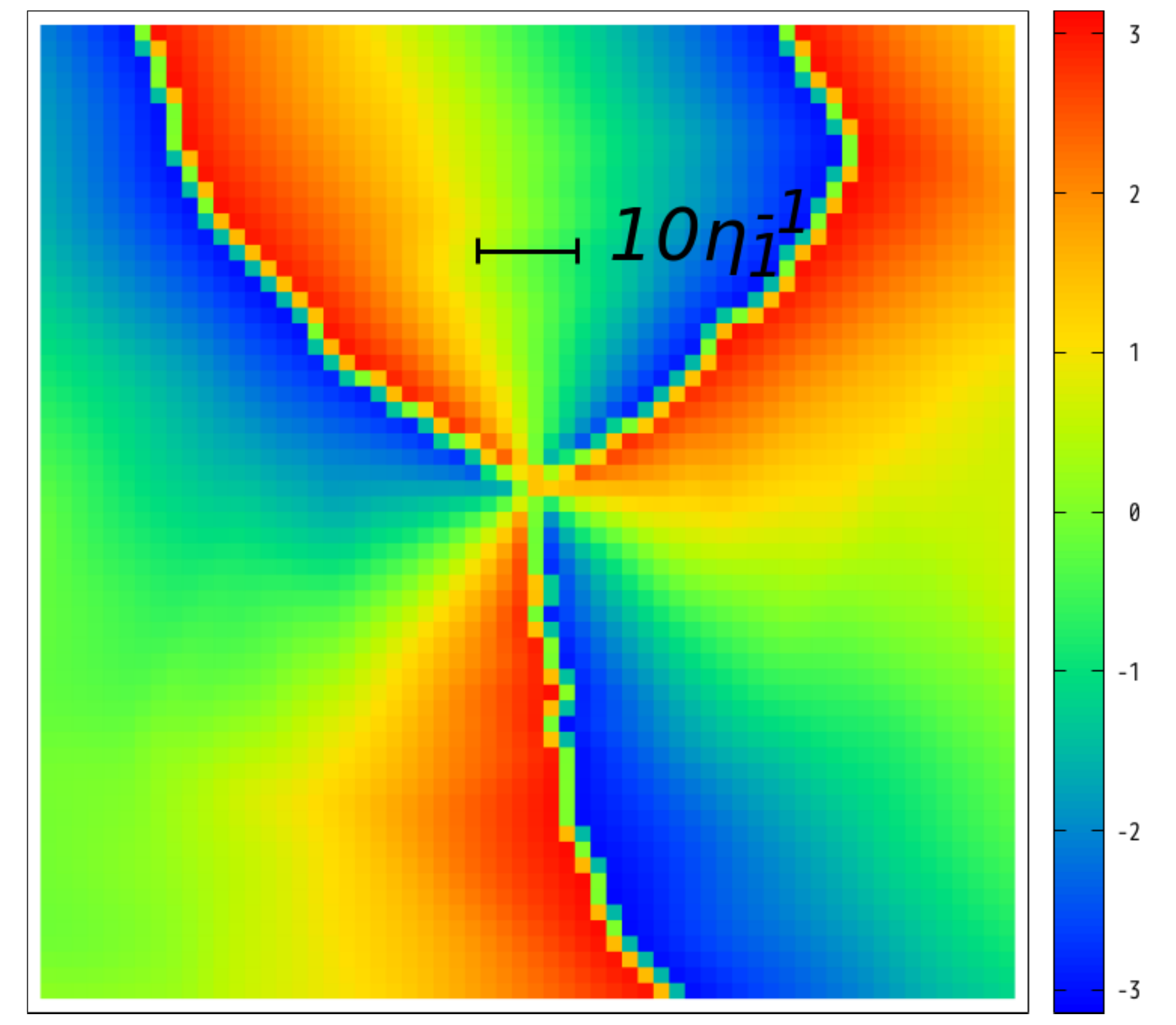}
\end{minipage}
\hspace{2mm}
\begin{minipage}{.175\textwidth}
\centering
\includegraphics[width=1.1\linewidth]{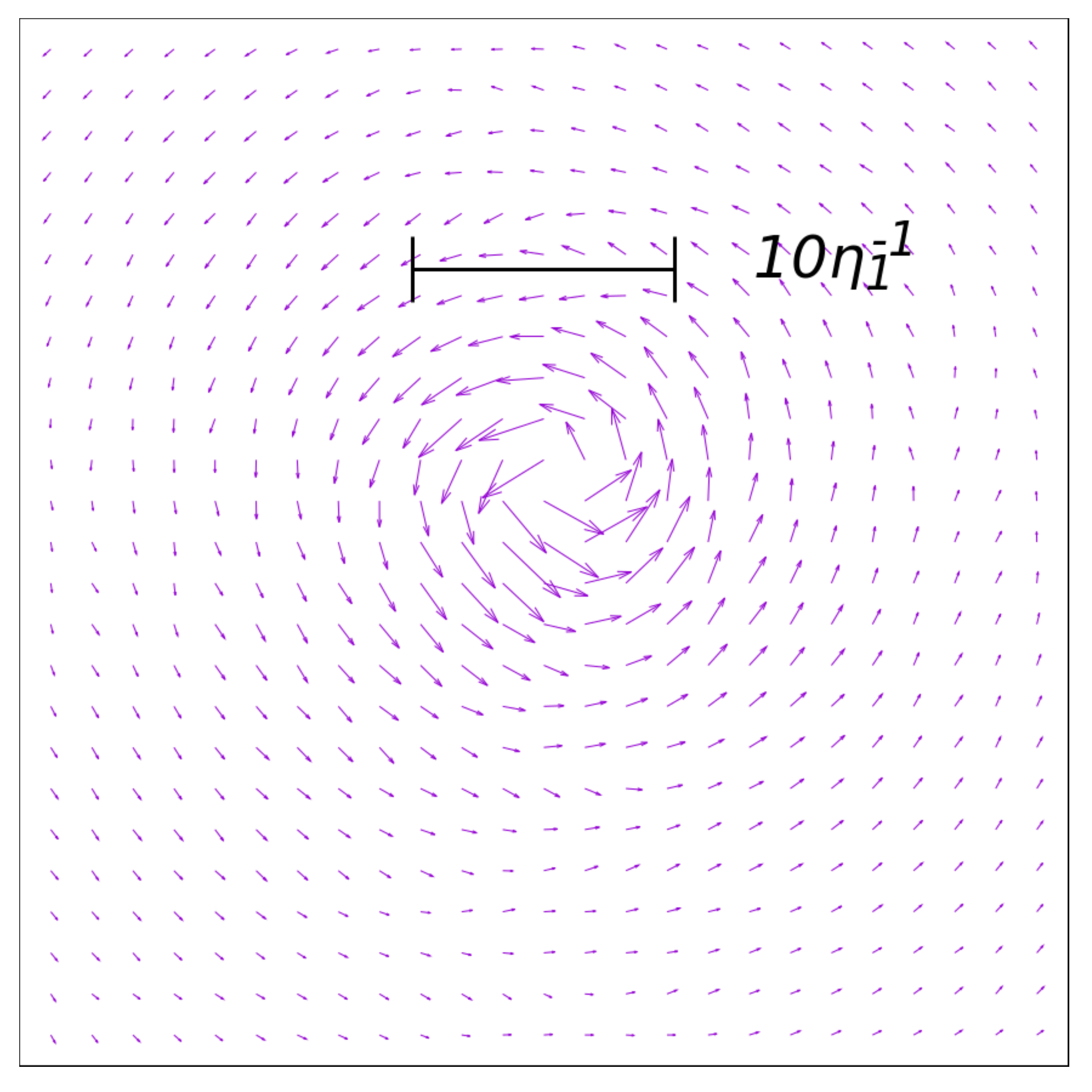}
\end{minipage}
\small{|| Spot 3 $(n_1,n_2) = (1,5)$ ||}\\
\begin{minipage}{.18\textwidth}
\centering
\includegraphics[width=1.2\textwidth]{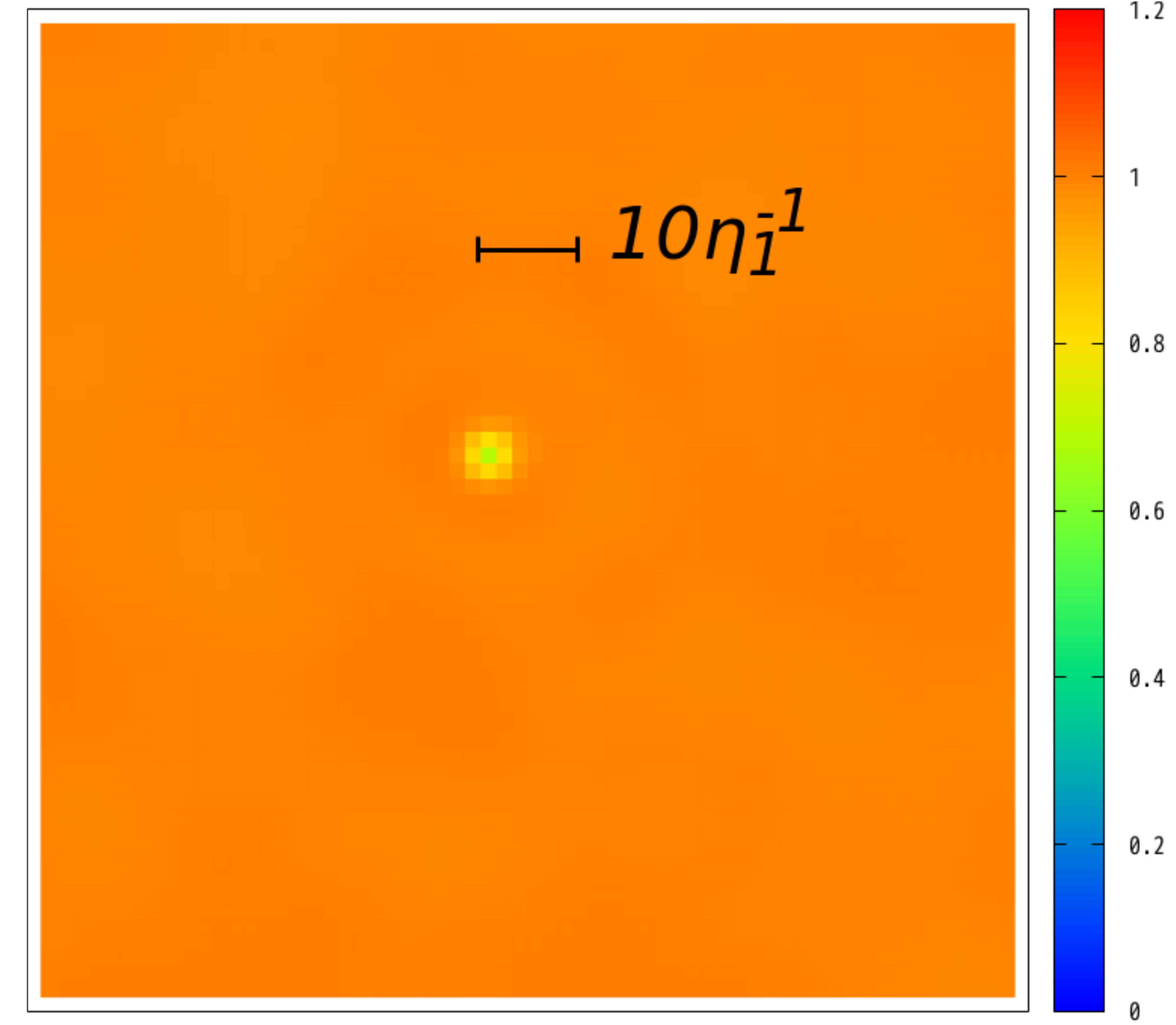} 
\end{minipage}
\begin{minipage}{.18\textwidth}
\centering
\includegraphics[width=1.2\linewidth]{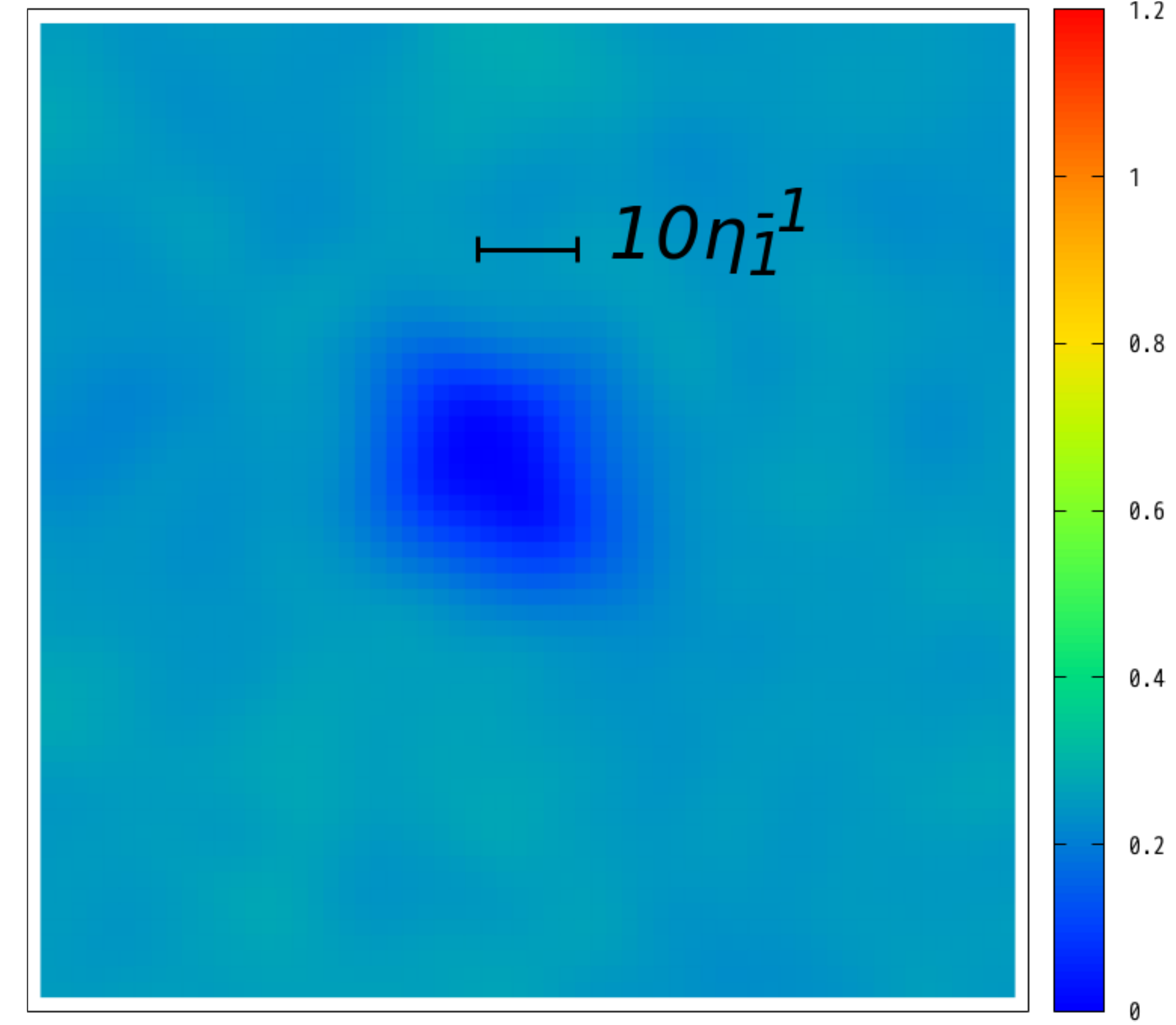}  
\end{minipage}
\hspace{2mm}
\begin{minipage}{.18\textwidth}
\centering
\includegraphics[width=1.2\linewidth]{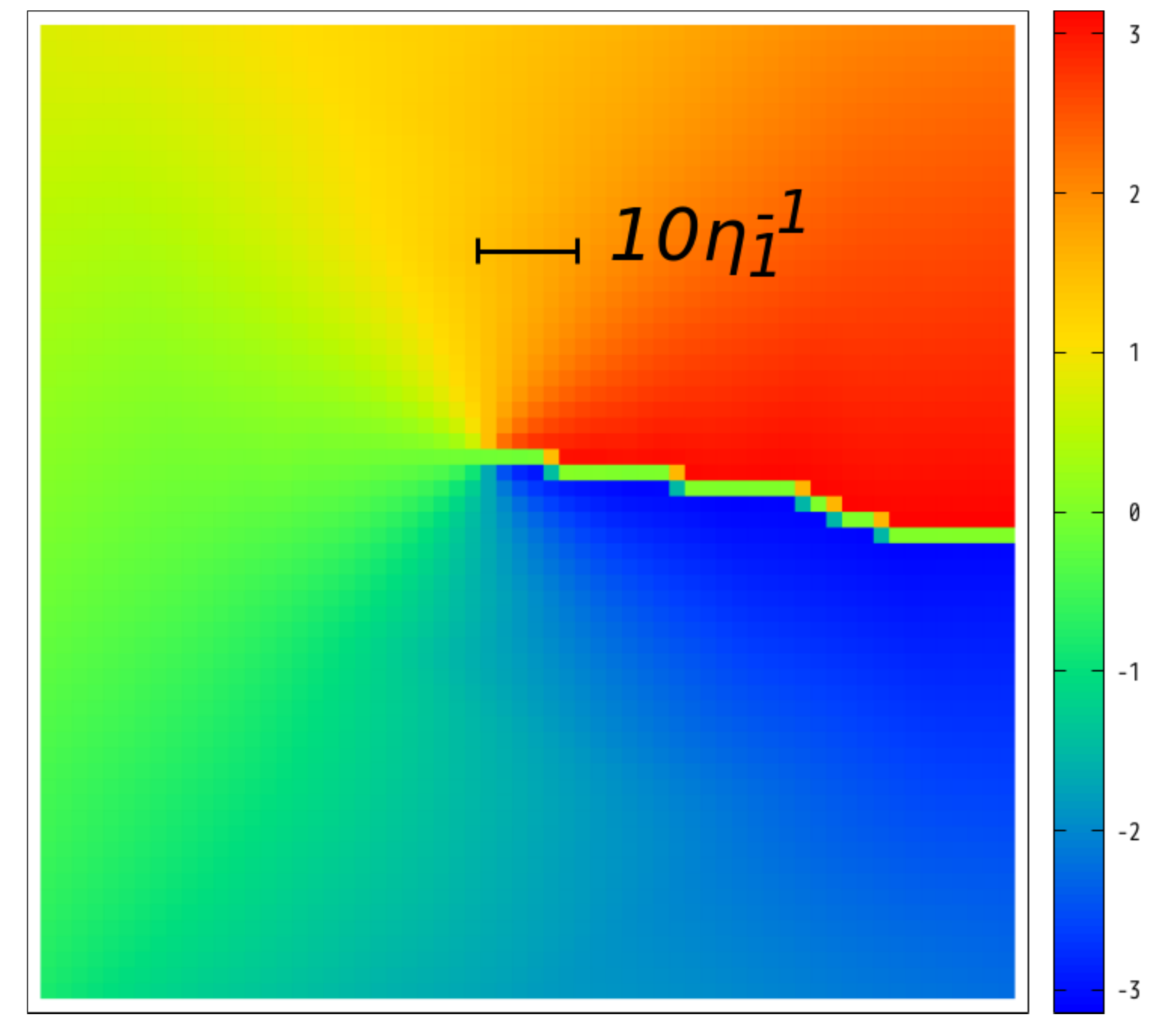}    
\end{minipage}
\begin{minipage}{.18\textwidth}
\centering
\includegraphics[width=1.2\linewidth]{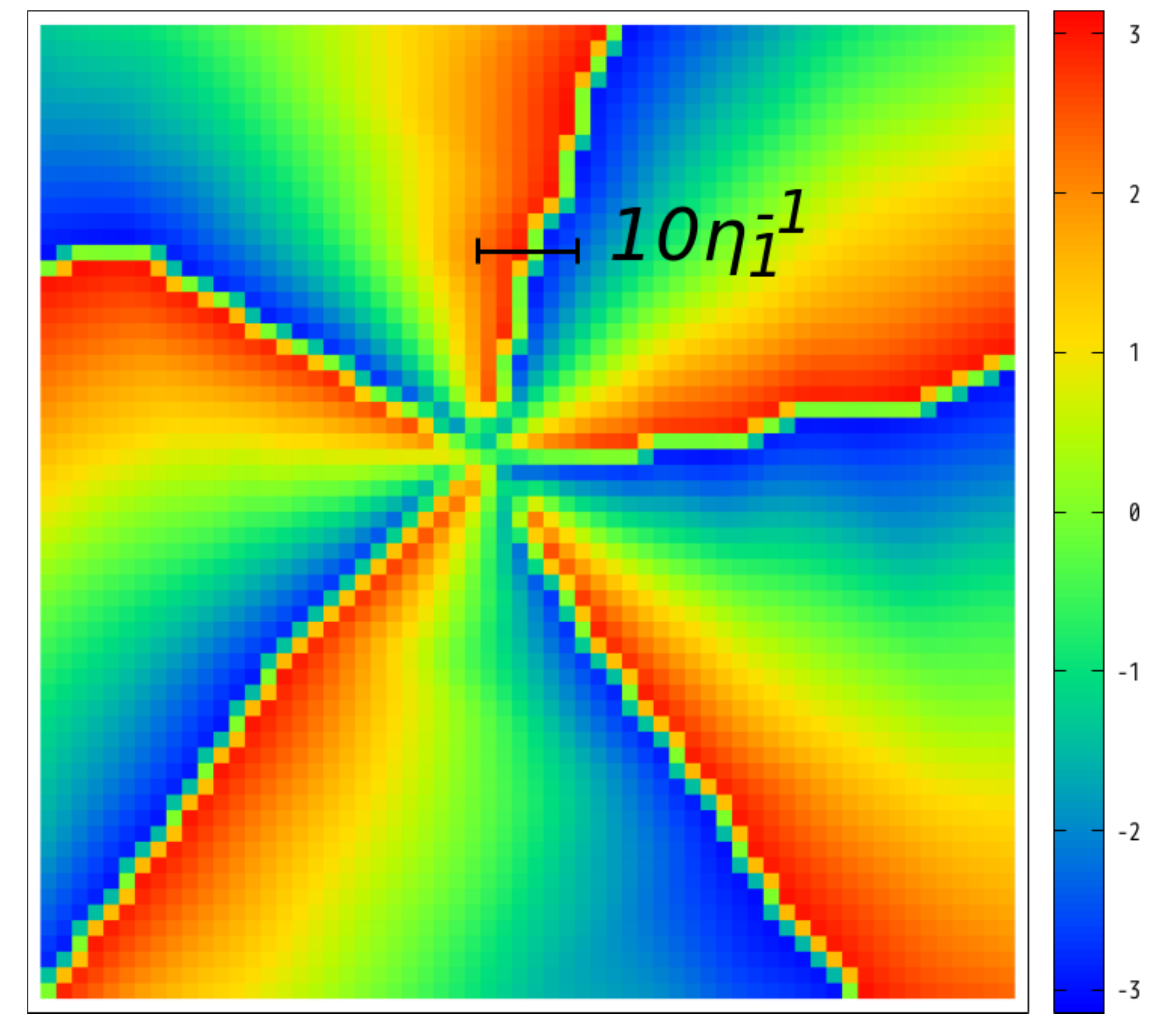}
\end{minipage}
\hspace{2mm}
\begin{minipage}{.175\textwidth}
\centering
\includegraphics[width=1.1\linewidth]{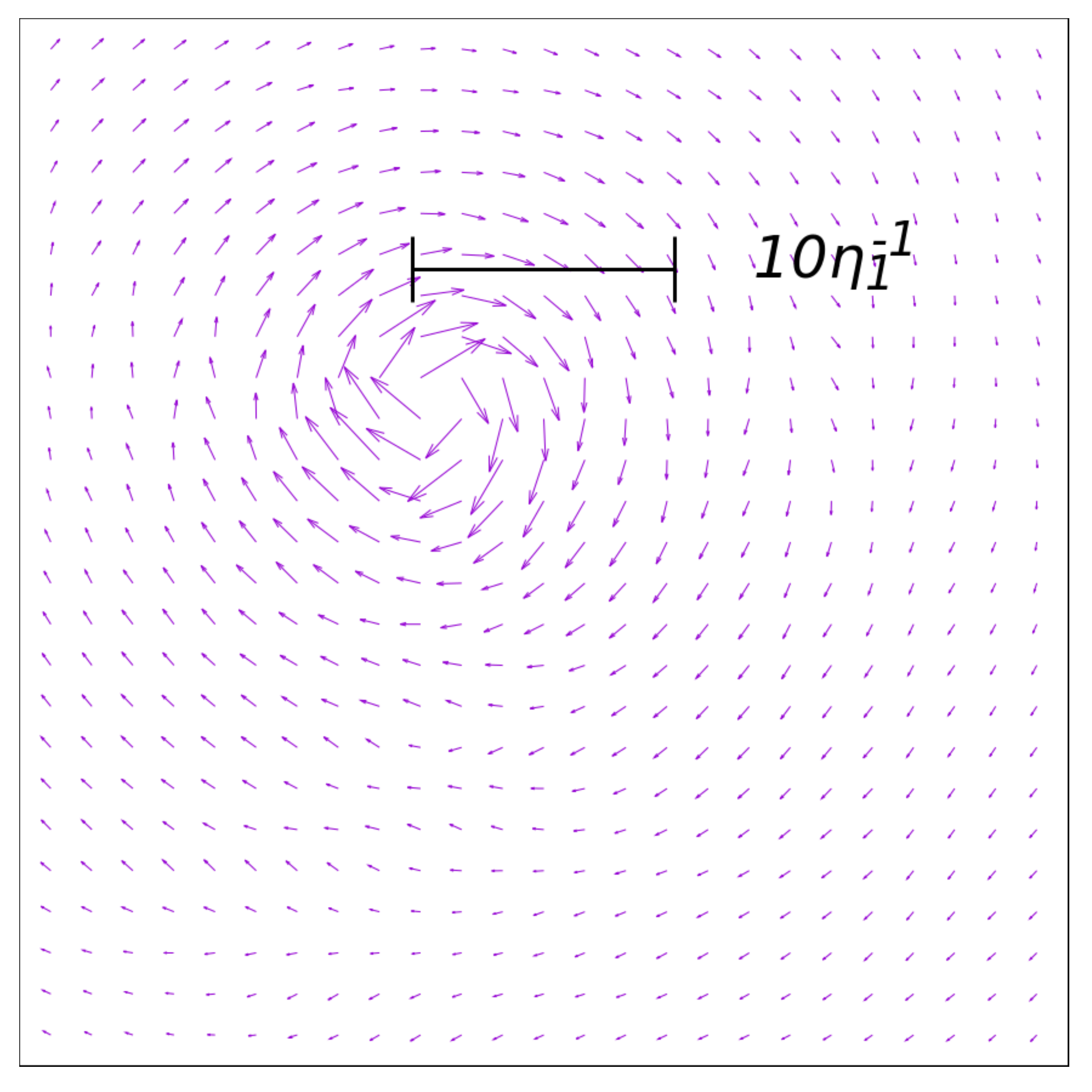}
\end{minipage}
\small{|| Spot 4  $(n_1,n_2) = (0,1)$ ||}\\
\begin{minipage}{.18\textwidth}
\centering
\includegraphics[width=1.2\textwidth]{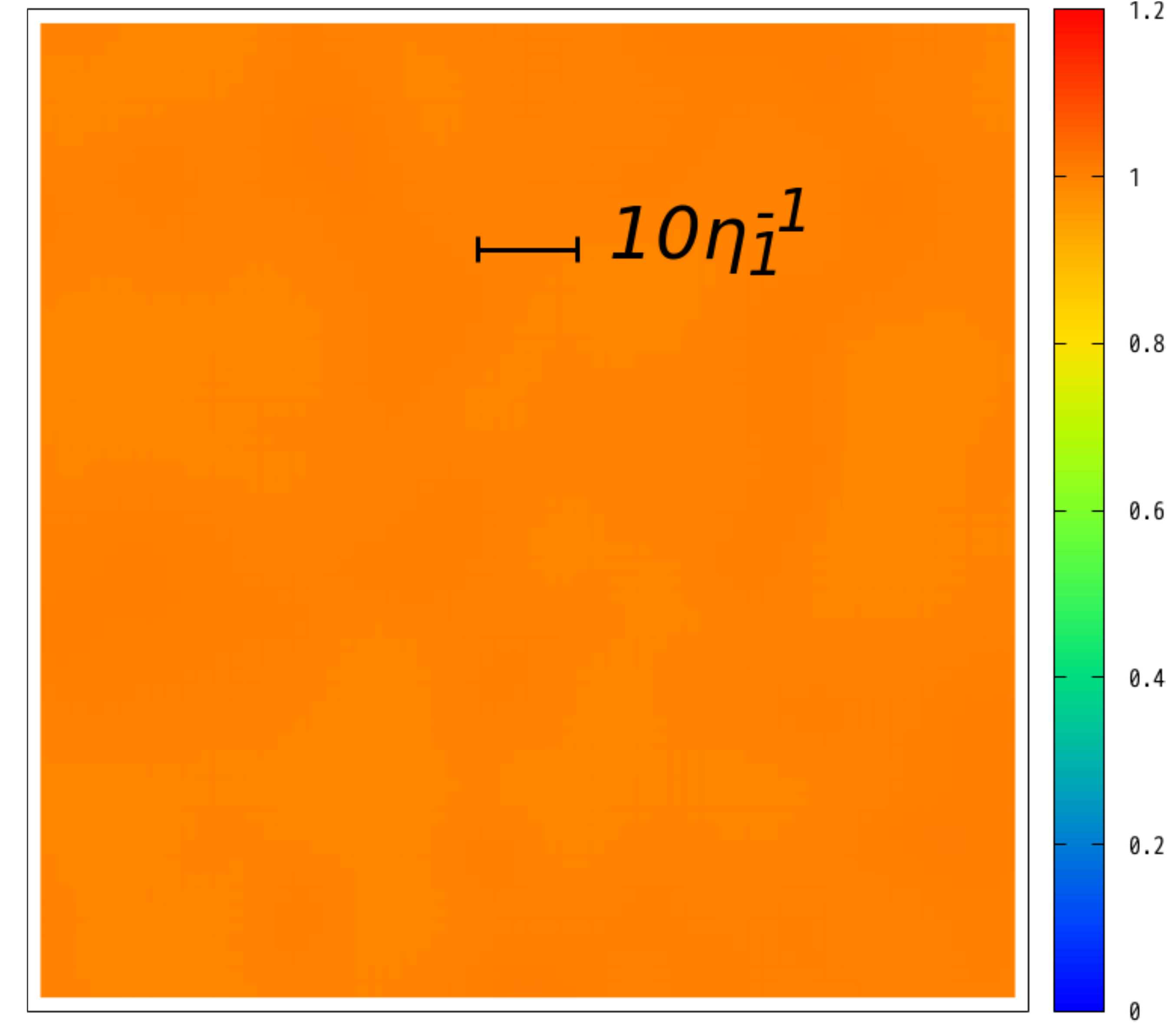} 
\end{minipage}
\begin{minipage}{.18\textwidth}
\centering
\includegraphics[width=1.2\linewidth]{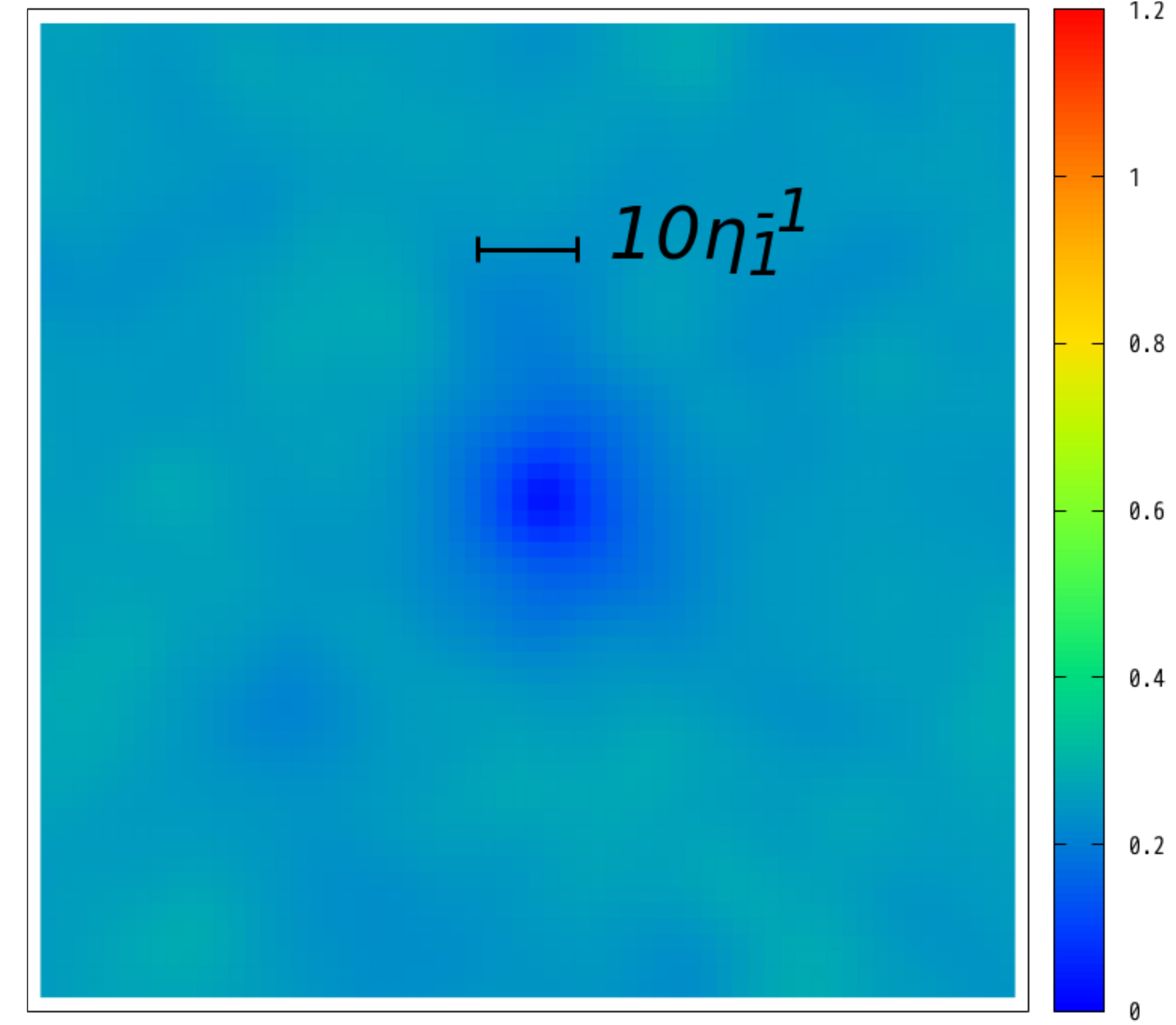}  
\end{minipage}
\hspace{2mm}
\begin{minipage}{.18\textwidth}
\centering
\includegraphics[width=1.2\linewidth]{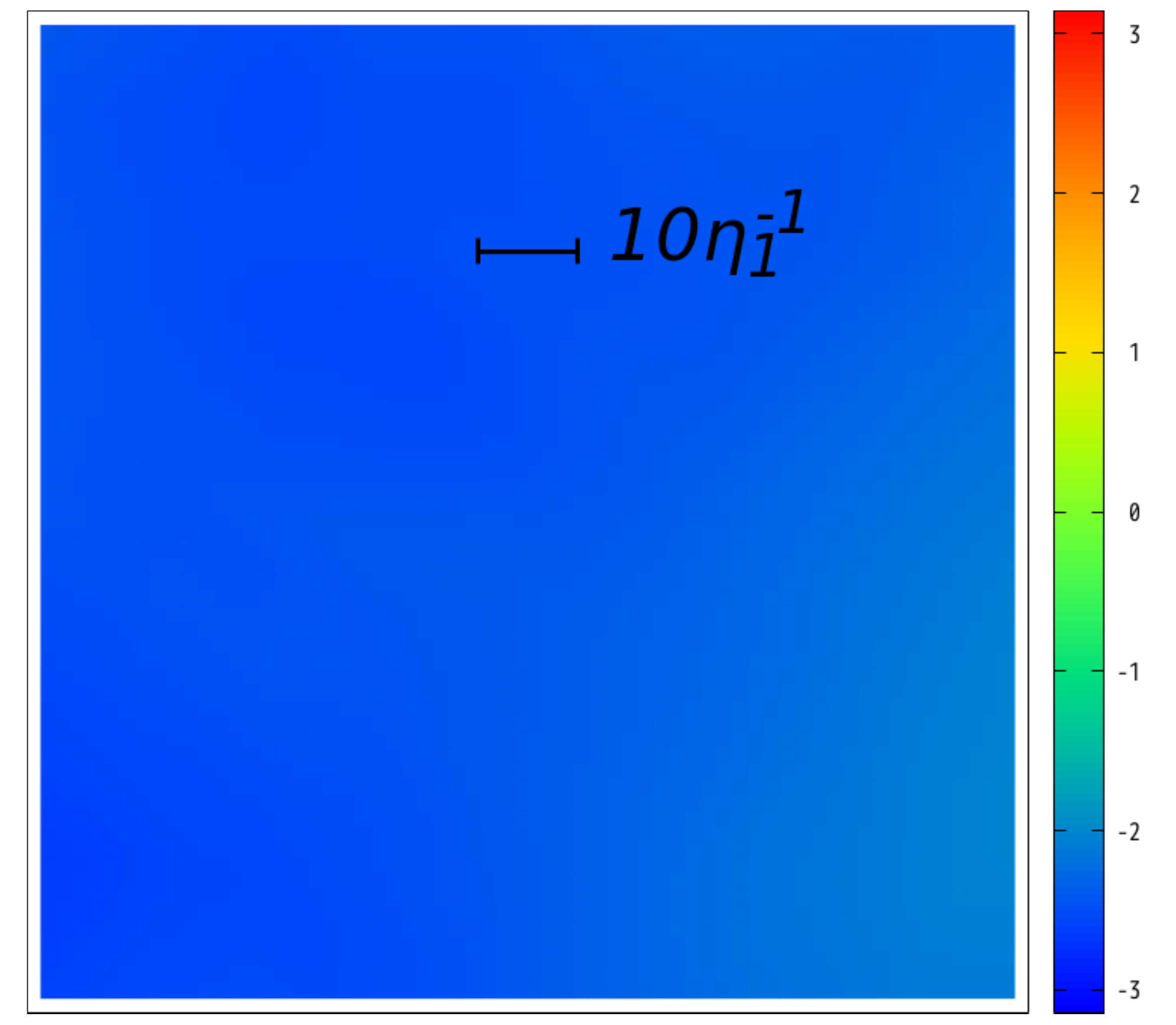}    
\end{minipage}
\begin{minipage}{.18\textwidth}
\centering
\includegraphics[width=1.2\linewidth]{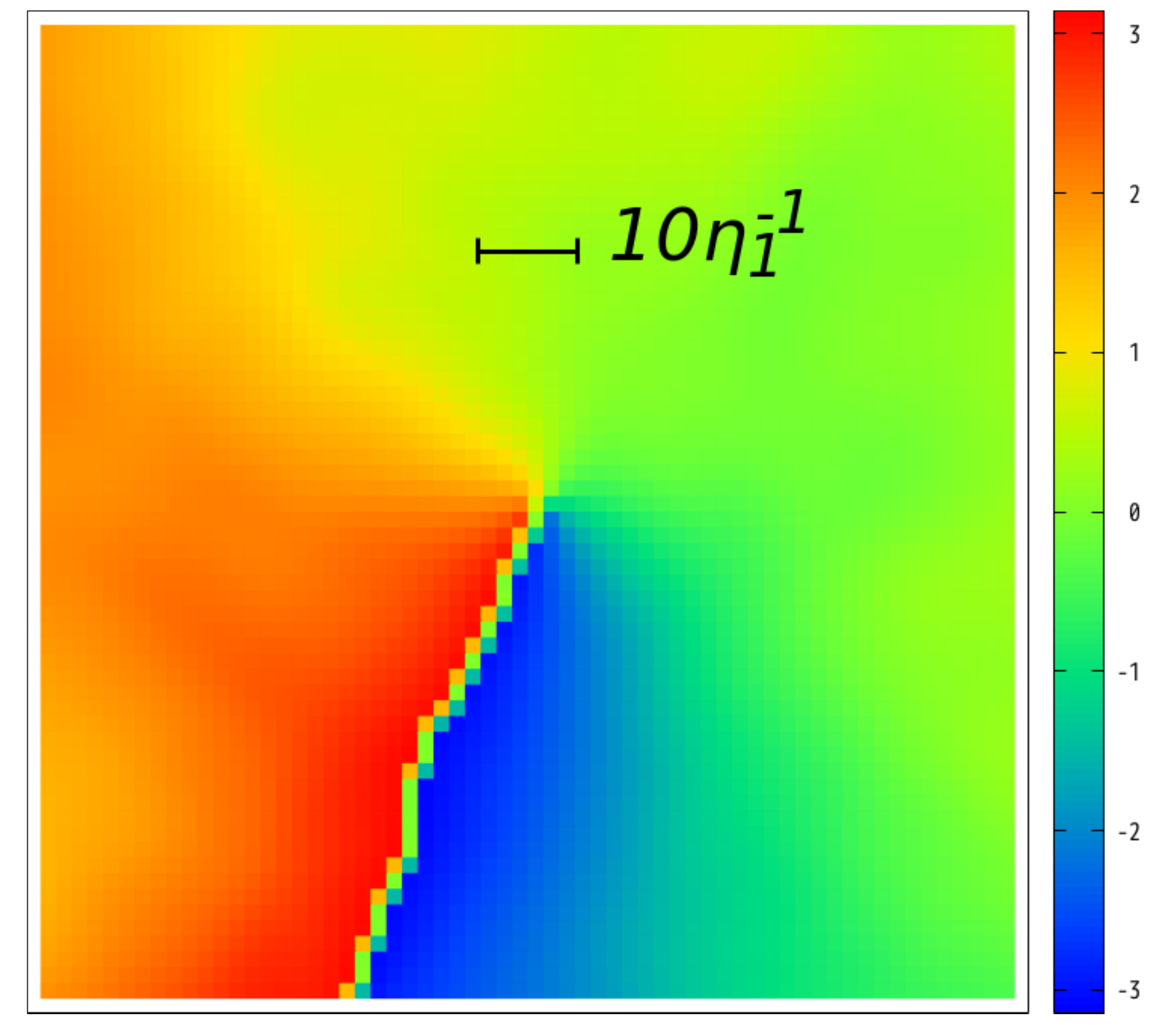}
\end{minipage}
\hspace{2mm}
\begin{minipage}{.175\textwidth}
\centering
\includegraphics[width=1.1\linewidth]{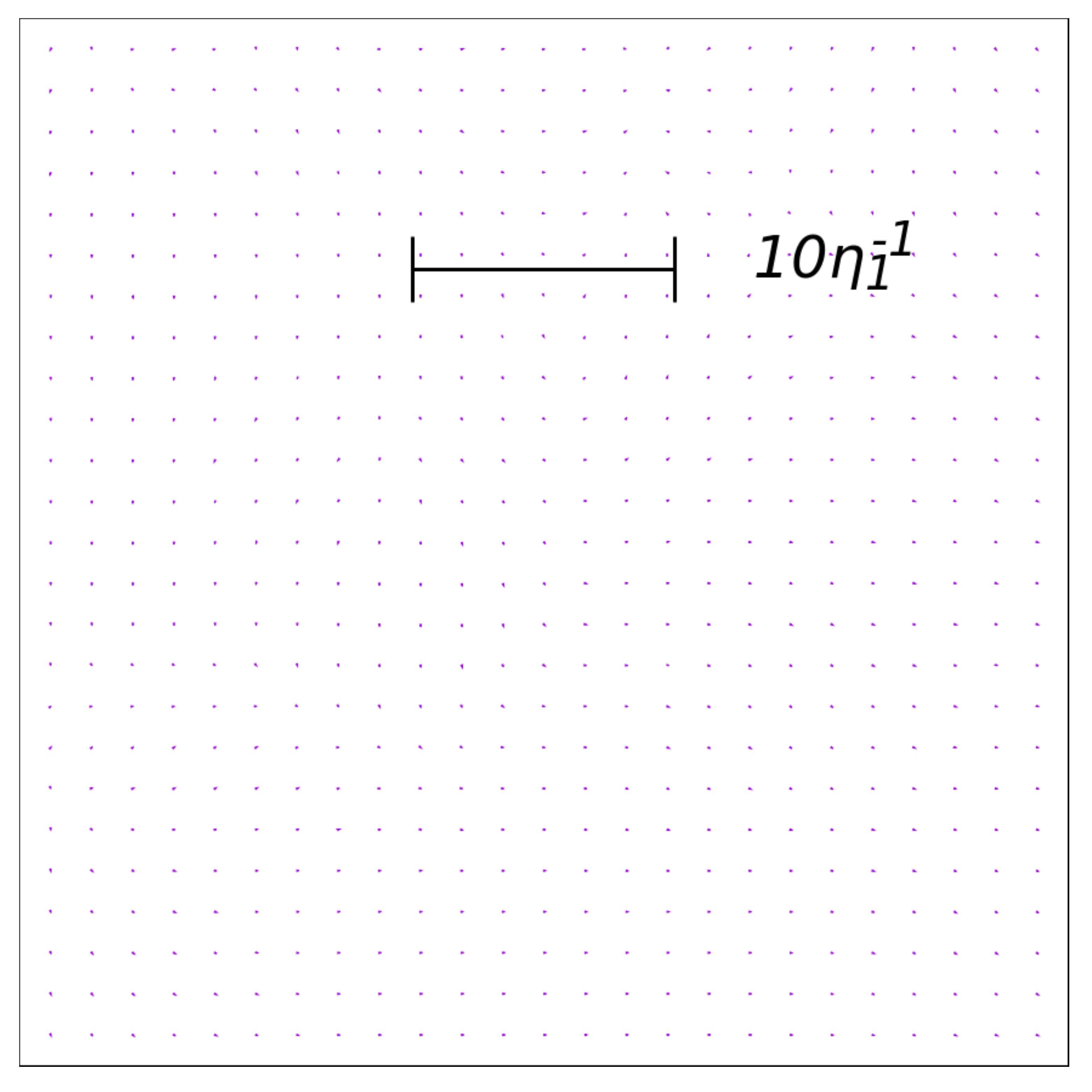}
\end{minipage}
\caption{
\sl \small\raggedright 
The field configurations in 
$2$-dimensional space after the phase transition.
The first row shows the field configuration in the whole two-dimensional simulation box.
The following rows are the closeups of the example spots with string configurations.
The right panels show the amplitude of the gauge field in the whole simulation box (Top),
and the gauge field in the $2$-dimensional space at each spot,
whose amplitude is proportional to the length of the vector.
}
\label{fig:Results}
\end{figure}

In Fig.~\ref{fig:Results}\footnote{Supplemental materials are available at {\tt http://numerus.sakura.ne.jp/research/open/NewString/}.}, we show 
the field configurations in the 
$2$-dimensional space at the 
temperature, $T \ll \eta_{1,2} $, 
where the phase transition of the $U(1)$
symmetry breaking has completed.
From the left, the figures show the absolute field values of $\phi_1$
and  $\phi_2$ {normalized by $\eta_1$}, the phase distributions of $\phi_1$
and $\phi_2$, and the gauge field configurations.
The first row shows the field configurations in the whole $2$-dimensional simulation box
at the simulation end.
At that time, the side length of the box is twice of the Hubble horizon length. 
The following rows are the closeups of the example spots with string configurations.
The winding number of each spot is shown in the label.

The string configuration at the spot $1$
has the winding numbers, $(n_1,n_2)=(1,4)$.
This  configuration corresponds to
the compensated (local) string (see Eq.\,\eqref{eq:compensated}).
As discussed in Sec.~\ref{sec:stringsolution}, 
the non-trivial configuration of the gauge field cancels the covariant derivatives 
of the both complex scalars.
The string configurations at the 
spots $2$ and $3$ are, on the other hand, the uncompensated strings. 
As emphasized in the previous section, 
they are the new-types of the string solutions in the present model.
In these configurations, the covariant derivatives of the scalar fields are not canceled although 
they have non-trivial configurations of the gauge field around them.
The string configuration at the spot $4$
has the winding number $(n_1,n_2) = (0,1)$.
This corresponds to the 
global string which appears in the phase 
transition of the global $U(1)_{global}$ symmetry
breaking.
The global strings are not accompanied by non-trivial configurations of the gauge field.

In the present setup, we take $\eta_1/\eta_2 = 4$, and hence, 
the phase transition of the $U(1)_{local}\times U(1)_{global}$ breaking 
takes place in stages.
At the first stage, the $U(1)_{local}$ breaking takes place at $T\sim \eta_1$.
At the second stage, the $U(1)_{global}$ breaking takes place
at $T\sim \eta_2$.
At the first stage, the string configurations with $n_1\neq 0$ (mostly $n_1 = 1$) are
formed in the similar manner to the original Abelian Higgs case.
Then, around the string configurations 
with $n_1\neq 1$, $\phi_2$ also takes 
non-trivial configurations with various $n_2$ at the second stage.
These configurations become
either the compensated and the uncompensated strings.
At the second stage, it is also possible
that $\phi_2$ takes non-trivial  configurations with $n_2\neq 0$ in the region where $\phi_1$ is trivial, i.e.,  $\langle \phi_1 \rangle = \eta_1$.
These configurations are the global strings, in which $\langle\phi_1\rangle = \eta_1$ is  
barely affected by the non-trivial configurations of  
$\phi_2$
due to the hierarchy $\eta_1 \gg \eta_2$.

\subsection{Evolution of String Network}
\label{sec:stringevolution}

\begin{figure}
\begin{minipage}{.32\textwidth}
\centering
\subcaption{\small Number of strings}
\includegraphics[width=\linewidth]{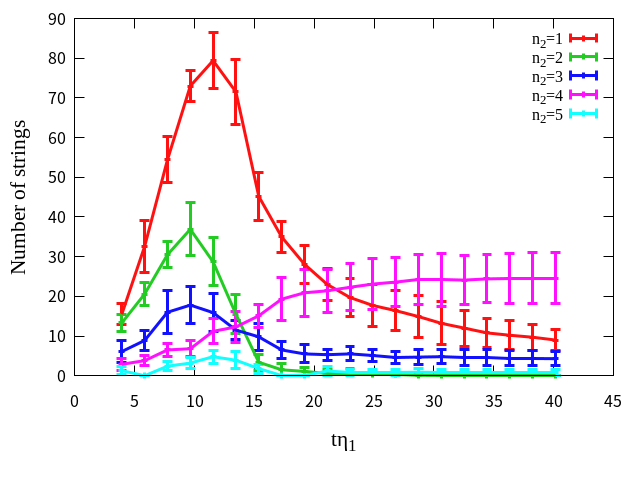} 
\end{minipage}
\begin{minipage}{.32\textwidth}
\centering
\subcaption{$f_{\rm c}, f_{\rm u}$ and $f_{\rm g}$}
\includegraphics[width=\linewidth]{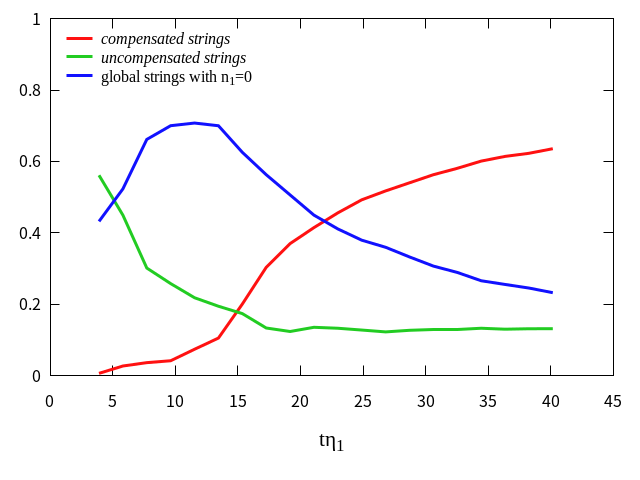}  
\end{minipage}
\vspace{2mm}
\begin{minipage}{.32\textwidth}
\centering
\subcaption{$R_{\rm c}$, $R_{\rm dw}$}
\includegraphics[width=\linewidth]{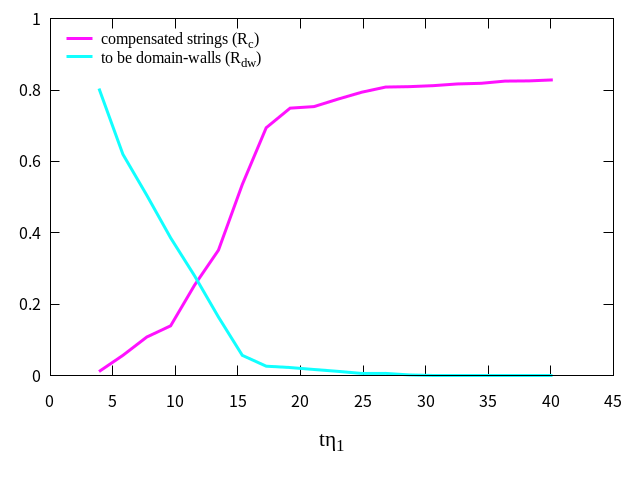}  
\end{minipage}
\caption{
\sl \small\raggedright 
{\it From left to right} : The time evolution of the numbers of the strings 
with various $n_2$ in the computational domain;
the number fractions of the compensated strings ($f_{\rm c}$),
the uncompensated strings ($f_{\rm u}$), and the global strings ($f_{\rm g}$);
and two kinds of number fractions ($R_{\rm c}, R_{\rm dw}$) 
defined in Eqs.\,\eqref{eq:Rc} and \eqref{eq:Rdw}.
The horizontal axis is the conformal time normalized by $\eta^{-1}_1$.
}
\label{fig:numofstrings}
\end{figure}
After the formation of the string configurations, the string network exhibits complicated evolution.
In Fig.~\ref{fig:numofstrings},
we show the time evolution of the 
numbers of the strings with various $n_2$ in the computational domain.
The results are obtained by averaging over 10 realizations, and 
the error bars show the statistical variance of the number of strings
arising from the randomness of the initial conditions.
To identify each string and compute its winding number for $\phi_1$ and $\phi_2$,
we use the algorithm developed by Ref.~\cite{Yamaguchi:2002sh}, and see also Ref.~\cite{Hiramatsu:2013tga}. Throughout this paper, we 
regard two strings separated within $20\eta_1^{-1}$ as one string.
We define $N_c$ as the number of compensated strings which have $(n_1,n_2)=(1,4)$,
$N_{\rm u}$ as that of uncompensated strings with $n_2 q_{1}-n_1 q_{2}\neq 0$ with  $n_1\ne 0$,
$N_{{\rm g},1}$ as that of global strings with $(n_1,n_2)=(0,1)$,
and 
$N_{{\rm g},>1}$ as that of global strings with $n_1=0$ and $n_2>1$.
The total number of string, $N_{\rm tot}$, is given as $N_{\rm tot}=N_{\rm c}+N_{\rm u}+N_{{\rm g},1}+N_{{\rm g},>1}$.
Note that in our simulations we observed no strings with $n_1>1$.
We also show the evolution of 
the number fractions of the compensated strings,
$f_{\rm c}=N_{\rm c}/N_{\rm tot}$ (red line),
the uncompensated strings,
$f_{\rm c}=N_{\rm u}/N_{\rm tot}$ (green line),
and the global strings, 
$f_{\rm g}=(N_{{\rm g},1}+N_{{\rm g},>1})/N_{\rm tot}$ (blue line) in the middle panels.
The right panels show the number ratio of the compensated strings to all the strings,
%
\begin{align}
    R_{\rm c} = \frac{N_{\rm c}}{N_{\rm c}+N_{\rm u}}\ ,
    \label{eq:Rc}
\end{align}
%
and the number ratio of the strings with $|n_2 q_{1}-n_1 q_{2}|>1$,
%
\begin{align}
    R_{\rm dw} = \frac{N_{|n_2 q_{1}-n_1 q_{2}|>1} }{N_{\mathrm{tot}}}\ .
    \label{eq:Rdw}
\end{align}
%
Here, $R_{\rm dw}$ includes contributions both from the uncompensated and the global strings with $|n_2 q_{1}-n_1 q_{2}|>1$.
This ratio is particularly important when the gauge-invariant Goldstone boson $a$ plays the role of the QCD axion.
In fact, the so-called the axion domain wall number around the cosmic strings are given by (a multiple of) $|n_2 q_{1}-n_1 q_{2}|$,
and the axion domain wall problem can be avoidable only for $|n_2 q_{1}-n_1 q_{2}| =0, 1$, if we 
assume that the phase transition of the $U(1)$  symmetries take place after the end of inflation (see the Appendix~\ref{sec:QCDaxion}).

The figure shows that the number fraction 
of the compensated local strings increases in time, 
while those of the uncompensated and the global strings 
decrease. 
This means that the uncompensated strings 
and the global strings tend to combine together to form the compensated strings.
After a while, the number of compensated strings tends to be converged, since the 
cosmic expansion separates the strings from each other and makes the compensation difficult.
In Sec.~\ref{sec:interaction},
we discuss how the long-range force 
between the uncompensated and the global strings
appears.

The simulation result in Fig.~\ref{fig:numofstrings}
also shows that the fraction
of the compensated strings converges to 
a value smaller than $1$, and hence, 
there remain uncompensated/global strings at later times.
We also found that $R_{\rm dw}\to 0$ in this simulation, 
which indicates that all the remaining strings in the network 
have the winding number either $0$ or 
$1$ in the gauge-invariant Goldstone direction.
This result has important implications for the 
axion model building (see the Appendix.~\ref{sec:QCDaxion}). 

\begin{figure}
\begin{minipage}{.32\textwidth}
\centering
\subcaption{\small Number of strings}
\includegraphics[width=\linewidth]{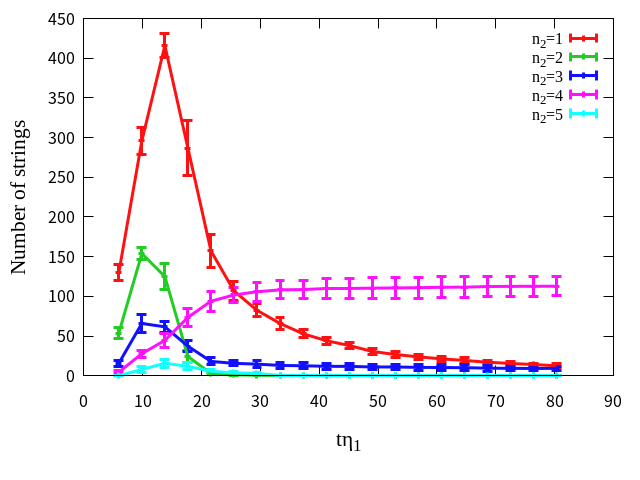}
\end{minipage}
\begin{minipage}{.32\textwidth}
\centering
\subcaption{$f_{\rm c}, f_{\rm u}$ and $f_{\rm g}$}
\includegraphics[width=\linewidth]{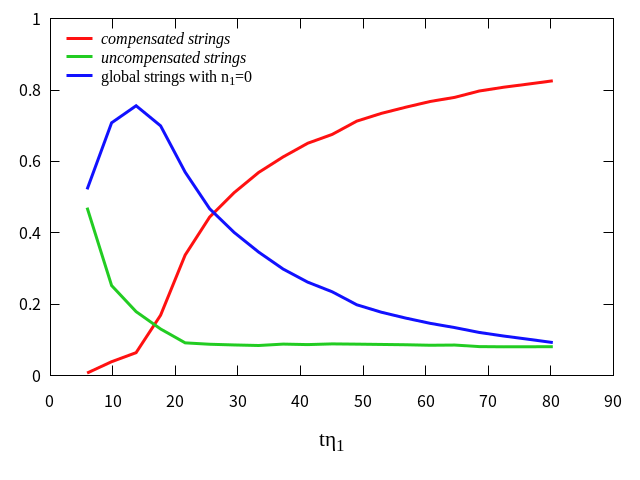}  
\end{minipage}
\vspace{2mm}
\begin{minipage}{.32\textwidth}
\centering
\subcaption{$R_{\rm c}$, $R_{\rm dw}$}
\includegraphics[width=\linewidth]{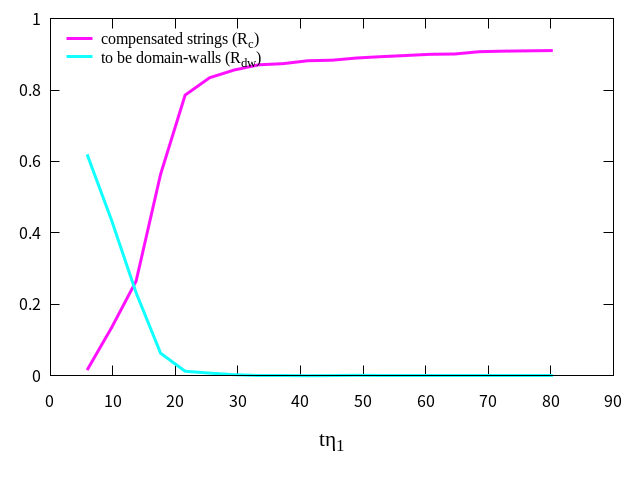}  
\end{minipage}
\caption{
\sl \small\raggedright 
The same figures as that shown in Fig.~\ref{fig:numofstrings} for 
the case including the thermal mass and $(q_{1},q_{2})=(1,4)$.
}
\label{fig:thermal}
\end{figure}

In the simulation of Fig.~\ref{fig:numofstrings},
we do not include the thermal mass term in Eq.\,\eqref{eq:thermalpot}.
Accordingly, the string formations of 
$\phi_1$ and $\phi_2$ take place simultaneously.
In a realistic situation, however, the string 
formation of $\phi_2$ takes place much later 
than that of $\phi_1$.
In Fig.~\ref{fig:thermal}, we show the same figures in 
Fig.~\ref{fig:numofstrings} with the effects of the thermal mass in Eq.~\eqref{eq:thermalpot}
taken into account.  
In this setup, the formation of the second string is delayed.
Hence we performed the simulations with a larger box whose initial size is given
by $80H_{\rm in}^{-1}$ and the grid size being $4096^2$.
Even in this case, the winding number of most of strings is $n_2=4$
and thus they are compensated. 
So we can conclude that the string network at late time is 
not sensitive to the presence of the thermal effect.
From now on, we do not take the thermal mass into account for the following simulations.

\begin{figure}[t]
\begin{minipage}{.32\textwidth}
\centering
\subcaption{\small Number of strings}
\end{minipage}
\begin{minipage}{.32\textwidth}
\centering
\subcaption{$f_{\rm c}, f_{\rm u}$ and $f_{\rm g}$}
\end{minipage}
\begin{minipage}{.32\textwidth}
\centering
\subcaption{$R_{\rm c}$, $R_{\rm dw}$}
\end{minipage}

\small{|| $(q_{1},q_{2}) = (1,1)$ ||}\\

\begin{minipage}{.32\textwidth}
\includegraphics[width=\linewidth]{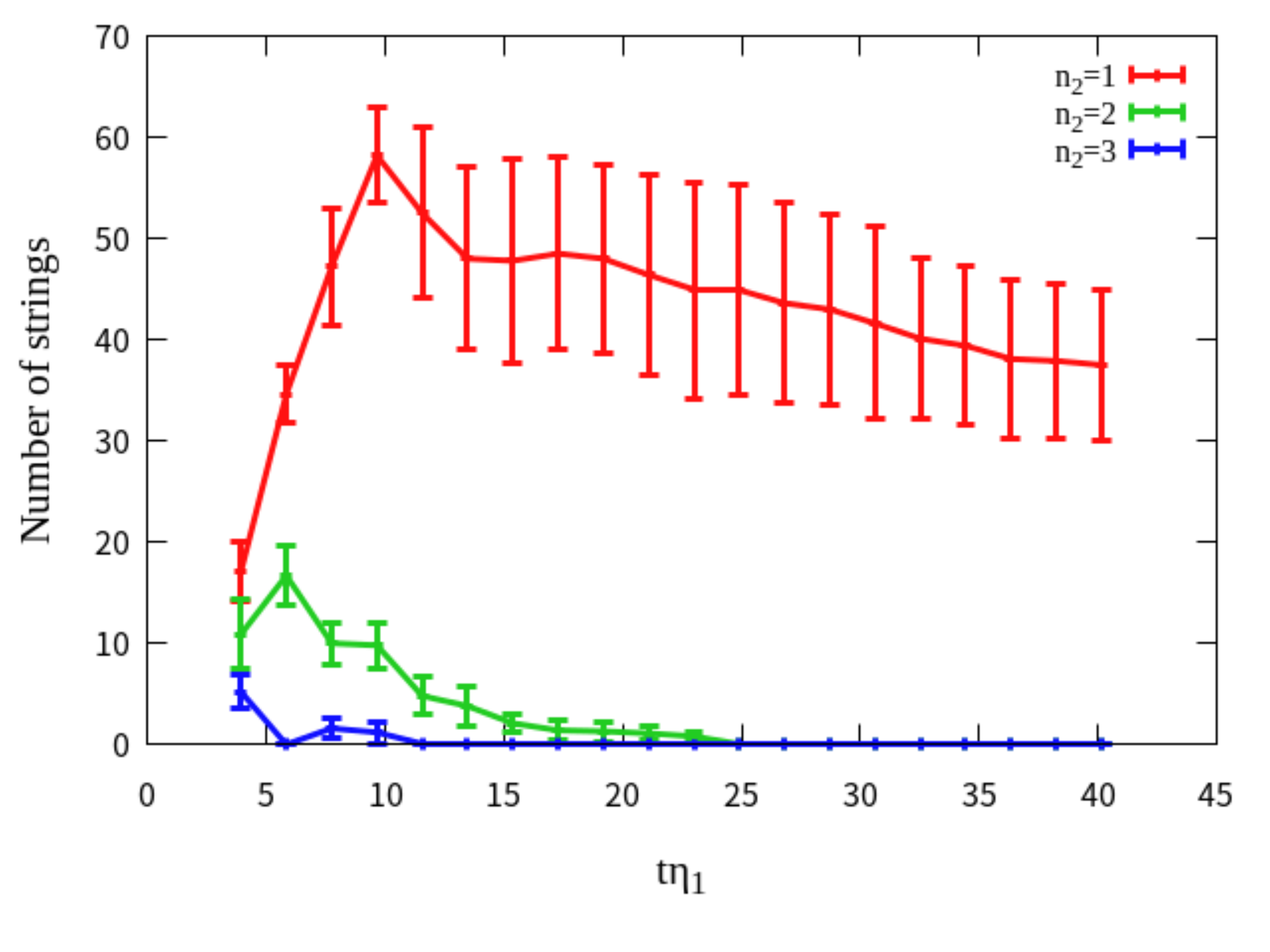} 
\end{minipage}
\begin{minipage}{.32\textwidth}
\centering
\includegraphics[width=\linewidth]{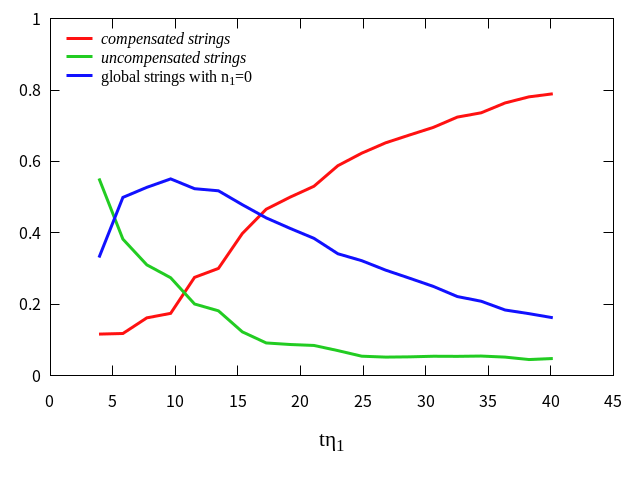}  
\end{minipage}
\begin{minipage}{.32\textwidth}
\centering
\includegraphics[width=\linewidth]{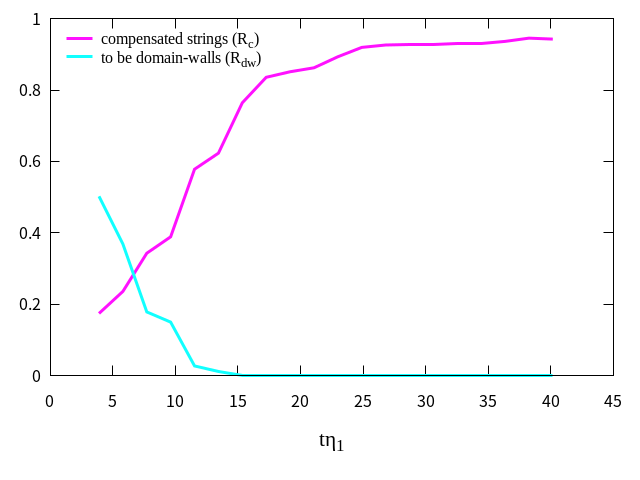}
\end{minipage}

\small{|| $(q_{1},q_{2}) = (1,2)$ ||}\\

\begin{minipage}{.32\textwidth}
\includegraphics[width=\linewidth]{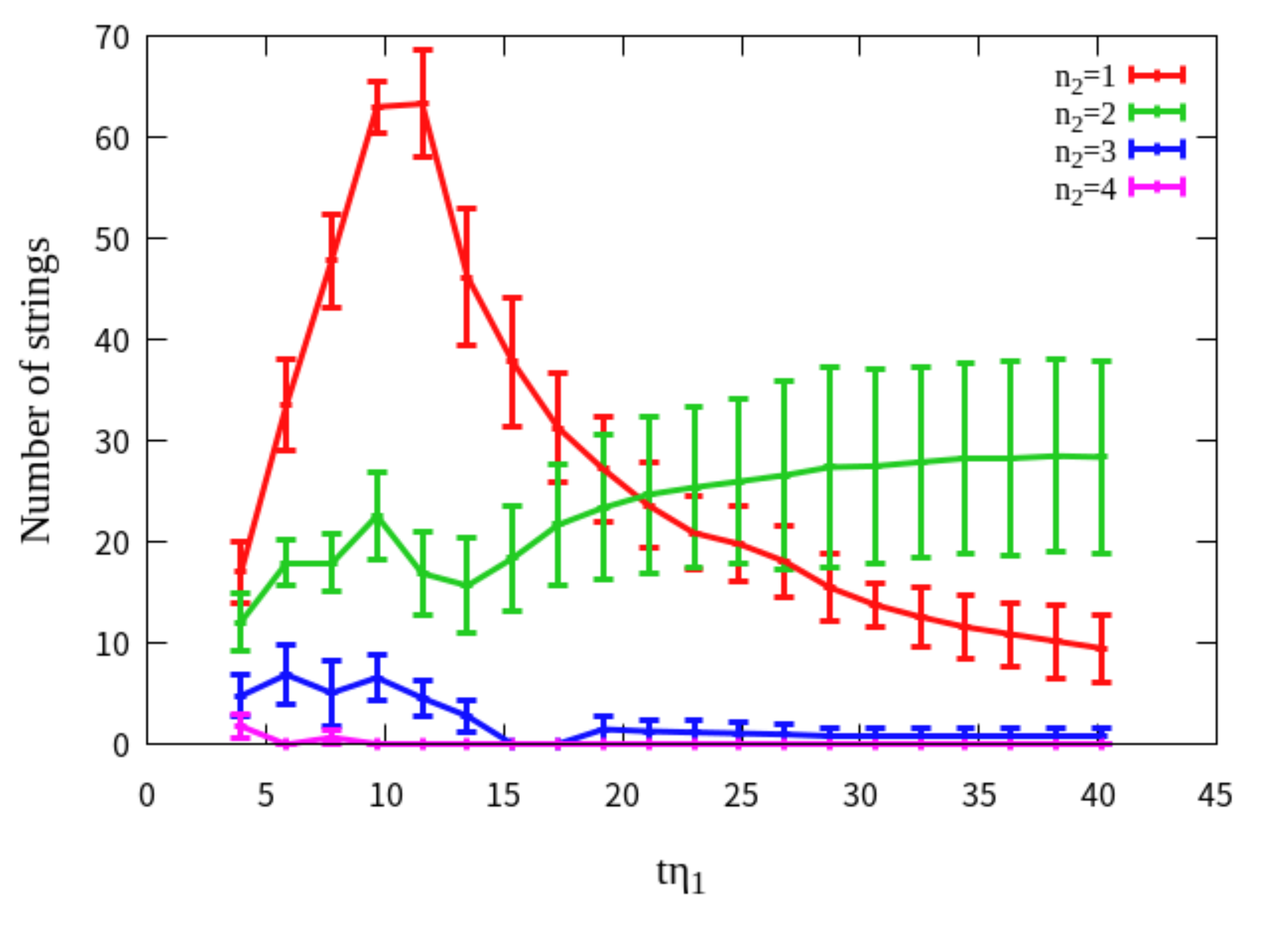} 
\end{minipage}
\begin{minipage}{.32\textwidth}
\centering
\includegraphics[width=\linewidth]{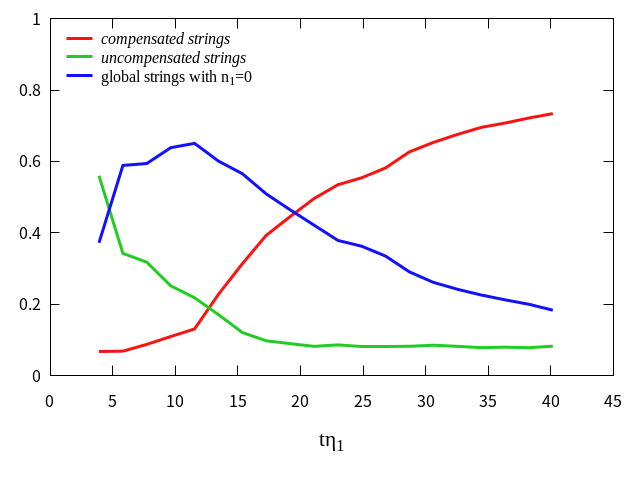}  
\end{minipage}
\begin{minipage}{.32\textwidth}
\centering
\includegraphics[width=\linewidth]{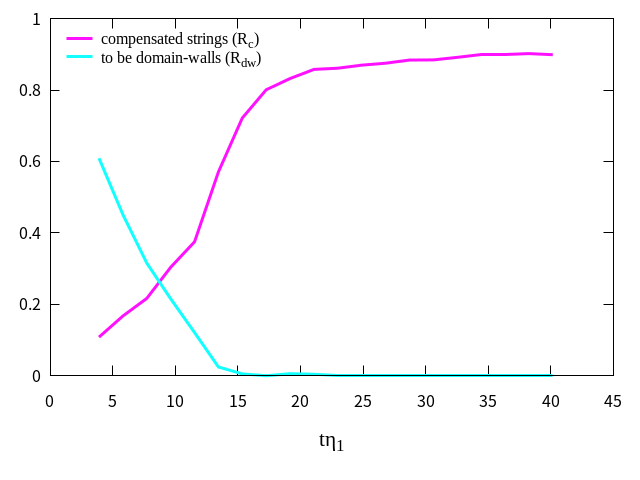}
\end{minipage}

\small{|| $(q_{1},q_{2}) = (4,1)$ ||}\\

\begin{minipage}{.32\textwidth}
\includegraphics[width=\linewidth]{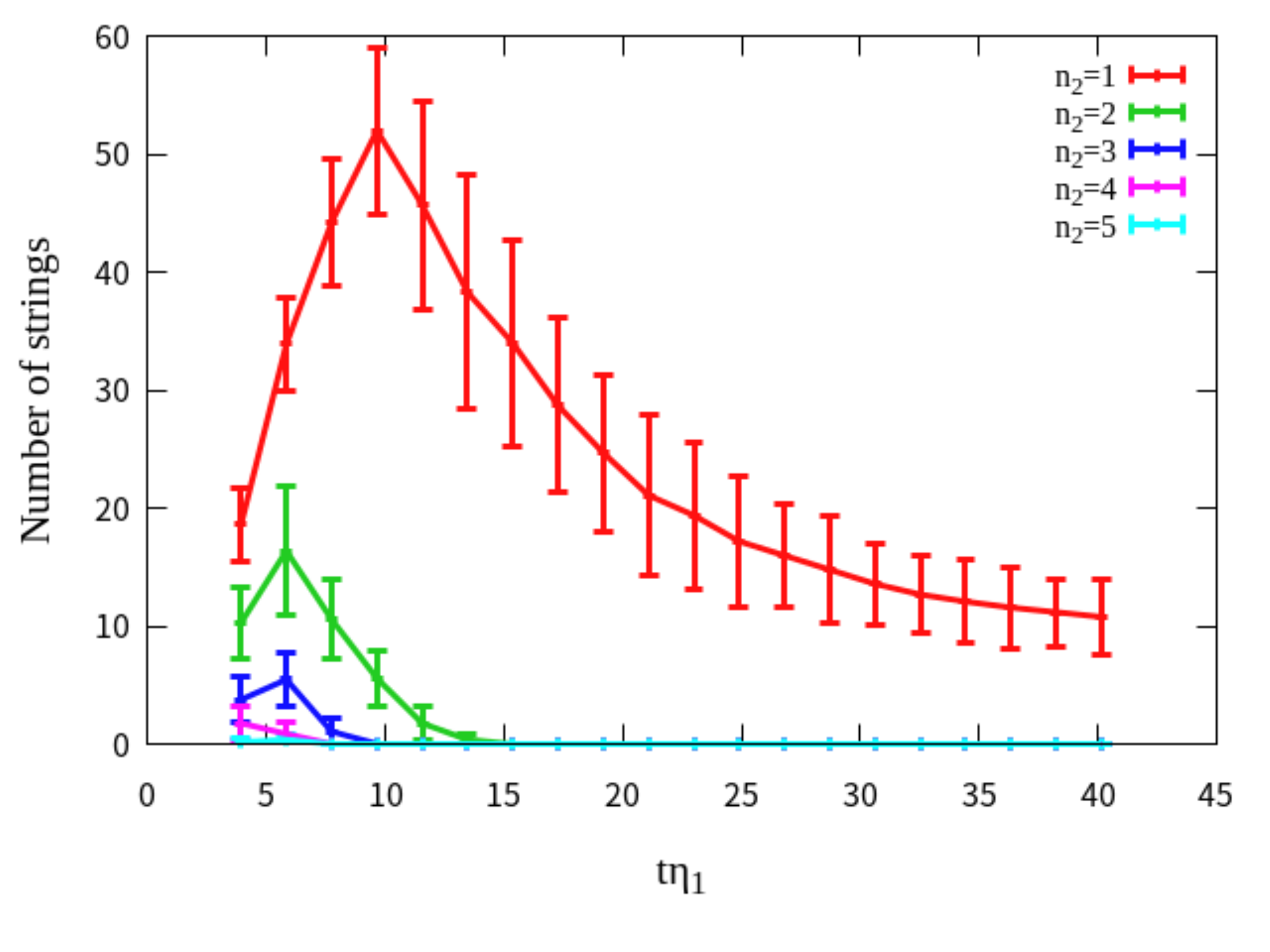} 
\end{minipage}
\begin{minipage}{.32\textwidth}
\centering
\includegraphics[width=\linewidth]{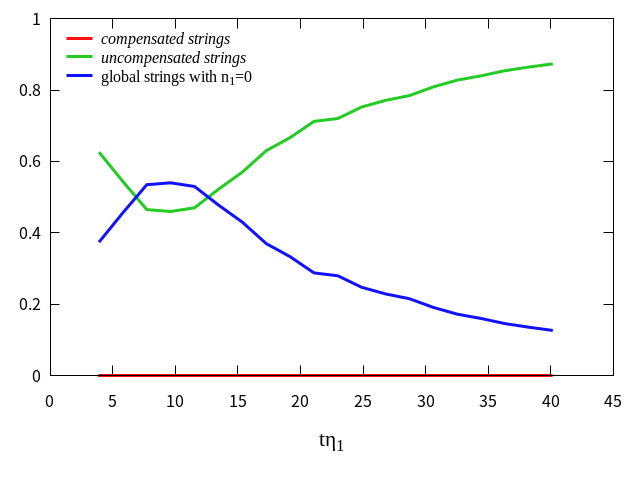}  
\end{minipage}
\begin{minipage}{.32\textwidth}
\centering
\includegraphics[width=\linewidth]{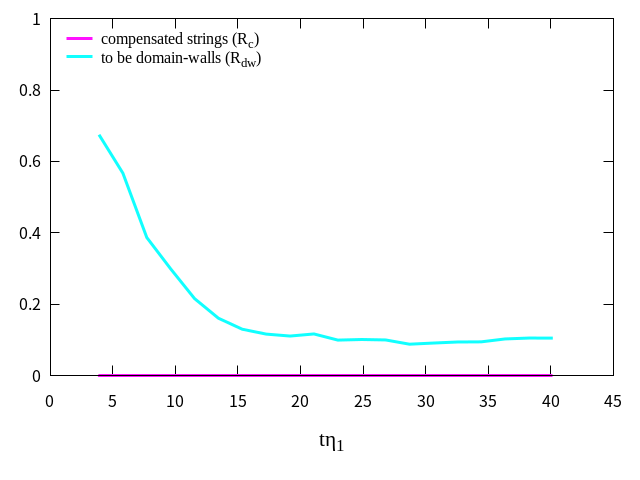}
\end{minipage}

\caption{
\sl \small\raggedright
The same figures with Fig.~\ref{fig:numofstrings} but for $(q_{1},q_{2})=(1,1), (1,2)$ and $(4,1)$.
}
\label{fig:numofstrings_others}
\end{figure}

\subsection{Parameter Dependence}

So far, we have fixed the model parameters to the values in Eq.\,\eqref{eq:parameters}.
Here, we briefly discuss parameter dependencies of the 
behavior of the string network.

To see the dependence on the charge ratio,
we have performed numerical simulations 
for $(q_{1},q_{2})=(1,1)$ and $(1,2)$,
while the other parameters are fixed.
In these cases, we have confirmed that 
the string networks show similar behavior to the case of $(q_{1},q_{2}) = (1,4)$,
and the number ratio of the compensated strings 
increases in time as shown in Fig.~\ref{fig:numofstrings_others}.
In fact, the compensated strings are dominated in the late time and thus most of strings 
take $n_2=q_2$.

We have also performed numerical simulations 
for $(q_{1},q_{2}) = (4,1)$ with 
the other parameters fixed.
In this case, we found that the string configurations have either $(n_{1}, n_{2}) = (1, 0)$ or $(n_{1}, n_{2}) = (0, 1)$ even at late times, and they do not combine into the compensated strings
unlike the case of $(q_{1},q_{2}) = (1,4)$ as shown in the bottom panels of Fig.~\ref{fig:numofstrings_others}.
This can be understood by the fact that 
the long-range force between the uncompensated 
and the global strings are proportional to the 
covariant derivative of $\phi_2$ around 
the uncompensated strings (see Sec.\,\ref{sec:interaction}).
In fact, ${\mathscr{D}}_\theta\phi_2$ around the string configurations of $(n_1,n_2) = (1,0)$ is smaller for $(q_{1},q_{2}) = (4,1)$ than that for $(q_{1},q_{2}) = (1,4)$.
As a result, we find that the string network consists of the strings with $(n_1,n_2) = (1,0)$ and $(n_1,n_2) = (0,1)$ for $(q_{1}, q_{2}) = (4,1)$
even at a later time.

Altogether, our lattice simulations in the $2+1$ dimensional spacetime suggest that the string network 
is dominated by the compensated strings $R_c \lesssim 1$ for $q_{1} < q_{2}$ and $\eta_1>\eta_2$,
while it is dominated by the uncompensated one  for $q_{1} > q_{2}$ and $\eta_1>\eta_2$.
We also find that $R_{\rm dw} \to 0$  for $q_{1} < q_{2}$ and $\eta_1>\eta_2$,
We will discuss the behavior of the string network in the $3+1$ dimensional spacetime in a separate paper,
where the $R_c$ and $R_{\rm dw}$ can be drastically different from those in the present simulations,
since the reconnection rates of the strings can be drastically different in  the $3+1$ dimensional spacetime. 

We have also performed the simulation for $\kappa=0.3$. 
In this case, the mass term of $\phi_2$ becomes positive at around $\phi_1 = 0$ which stabilizes the symmetry enhancement point, $\phi_2=0$ 
at the center of string core of $\phi_1$.
This feature is expected to make the correlation between $\phi_1$ and $\phi_2$ stronger.
We have numerically confirmed that the number of $\phi_2$ strings formed around the $\phi_1$ string tends to be larger for $\kappa>0$ 
than the case with $\kappa=0$,
although we do not discuss details of those tendencies in this paper 
(see also \cite{Urrestilla:2007yw}).

\section{Interaction between Strings}
\label{sec:interaction}

In this section, we discuss a long-range force 
between a compensated/uncompensated string 
with $(n_{1},n_{2})= (n_{1}\neq 0,n_{2A})$
and a global string with 
$(n_1,n_2) = (0,n_{2B})$.
We call the former to be the string A 
and the latter to be the string B.
As our simulation was performed in 
$2+1$ dimensional spacetime, 
we also consider the string configurations
in the $2$-dimensional space. 
In the $2$-dimensional space, 
the spatial coordinates of the centers of the string A and B are given by
${\mathbf{r}_A}$ and $\mathbf{r}_B$, respectively.

In our discussion, we assume $\eta_1\gg \eta_2$.%
\footnote{We also assume $\kappa = 0$.}
In this limit, the configurations of $\phi_1$ 
and the gauge field are hardly affected by the configuration of $\phi_2$, and hence, 
we can treat them as the fixed background fields.
Besides, we may neglect the $\phi_2$ contribution
to the Euler-Lagrange equation of the gauge field in Eq.\,\eqref{eq:elgauge} as $c_1 (\simeq 1) \gg c_2$.
With such a gauge field configuration, $\mathscr{D}_\theta \phi_1$ vanishes 
at the large distance from the string A 
as in the case of the ANO string.

When the string A and B are well separated, the configuration of $\phi_2$
can be approximated by,
%
\begin{align}
\label{eq:ansatzint}
    \phi_{2AB}(\mathbf{r})
    \equiv 
    \phi_2(\mathbf{r};\mathbf{r}_A,\mathbf{r}_B) = 
    \frac{
    \phi_{2A}(\mathbf{r}-\mathbf{r}_A)
    \phi_{2B}(\mathbf{r}-\mathbf{r}_B)}
    {\eta_2}\ ,
\end{align}
%
where $\phi_{2A}$ and $\phi_{2B}$ 
are the isolated string configurations of the string A and B.
The long-range force  between the string A and B
can be read off from the difference of the string energy per unit length~\cite{Bettencourt:1994kf}, 
%
\begin{align}
    {\mit{\Delta}}E(|\mathbf{r}_A-\mathbf{r}_B|)= E(\{\phi_{2AB}\})-E(\{\phi_{2A}\})-E(\{\phi_{2B}\})\ .
\end{align}
%
where the energy functional is given by
%
\begin{align}
\label{eq:energyint}
E(\{\phi_2\})=\int d^2\mathbf{r}
\left[\left|\partial_r \phi_2\right|^2+
\frac{1}{r^2}\left|
{\partial_\theta} \phi_2-
ieq_{2}{A_\theta}\phi_2\right|^2+\frac{\lambda_2}{4}\left(|\phi_2|^2-\eta^{2}_2\right)^2\right]\ .
\end{align}
%
In this expression, 
we have taken ${\mathbf r}_A = 0$,
and $(r,  \theta)$ are the radius and the azimuthal angle around the string A. 
In this coordinate system, $A_\theta$ is the gauge field configuration around the string.

When the string A and B are well separated, ${\mit{\Delta}E}$ is approximately given by 
%
\begin{align}
    {\mit{\Delta}}E \simeq \int d^{2}{\mathbf{r}} \frac{1}{\eta^{2}_2}
    \left(\phi_{2A}\phi^*_{2B}\partial_r \phi^{*}_{2A}\partial_r\phi_{2B} 
    +\frac{1}{r^2}\phi_{2A}\phi^{*}_{2B} \mathscr{D}_\theta \phi^{*}_{2A}\partial_\theta\phi_{2B} + h.c. \right) \ .
    \label{eq:DE}
\end{align}
%
Besides, by remembering that
%
\begin{align}
    |\phi_{2A}|,|\phi_{2B}| \to \eta_2 + \order{r^{-2}}\ ,
\end{align}
%
at the large distance, 
the contributions from the first term in Eq.\,\eqref{eq:DE}
is subdominant compared to those of the second term.
As a result, we find 
%
\begin{align}
     {\mit{\Delta}}E& \simeq \int d^{2}{\mathbf{r}} \frac{1}{\eta^{2}_2}\left(\frac{1}{r^2}\phi_{2A}\phi^{*}_{2B} \mathscr{D}_\theta \phi^{*}_{2A}\partial_\theta\phi_{2B} + h.c. \right) \ , 
     \\ 
      &\simeq 2 \, n_{2B} \left(n_{2A}-n_1
      \frac{q_{2}}{q_{1}}\right) \eta^{2}_2 \int d^{2}{\mathbf{r}} \frac{1}{r^2}
      \frac{\partial \theta_B}{\partial \theta}\ .
      \label{eq:DE2}
\end{align}
%
Here, $\theta_B$ is the azimuth angle around the string B, and its derivative with respect to $\theta$ 
 is given by
%
 \begin{align}
     \frac{\partial \theta_B}{\partial \theta} = \frac{r(r - |{\mathbf{r}_A -\mathbf{r}_B }|\cos\theta)}{|\mathbf{r}-\mathbf{r}_B |^2} \ .
 \end{align}
%
Therefore, the energy difference is given 
%
\begin{align}
{\mit{\Delta}}E(|\mathbf{r}_A -\mathbf{r}_B|)\simeq n_{2B} \left(n_{2A}-n_1
      \frac{q_{2}}{q_{1}}\right)
      \eta^{2}_2F(|\mathbf{r}_A -\mathbf{r}_B|),
      \label{eq:DE3}
\end{align}
%
where $F(x)$ is a decreasing function of $x$.

The long-range force between the uncompensated and the global strings is given by
the derivative of ${\mit {\Delta}}E$
with respect to $|\mathbf{r}_A-\mathbf{r}_B|$.%
\footnote{To avoid the logarithmic divergence of $F(x)$, we need to regularize the integration in Eq.\,\eqref{eq:DE2} 
at a large radius.
The long-range force is, on the other hand, free from the divergence.
}
Thus, we find the long-range force 
is attractive for
%
\begin{align}
n_{2B} \left(n_{2A}-n_1\frac{q_{2}}{q_{1}}\right) < 0\ ,
\end{align}
%
while it is repulsive for 
%
\begin{align}
n_{2B} \left(n_{2A}-n_1\frac{q_{2}}{q_{1}}\right) > 0\ .
\end{align}
%
With this force, the uncompensated 
string with $n_1= 1$ and $n_2 < 4$
and the global string with $n_1 = 0$
and $n_2 = 1$ combine together rather quickly for $(q_{1},q_{2})=(1,4)$.%
\footnote{The long-range attractive force presents even for $n_{2A} = 0$
where $h_{2}$ has a non-trivial configuration due to
the non-trivial $A_\theta$ configuration.}
The expression in Eq.\,\eqref{eq:DE3}
also explains the weakness of the 
attractive force between the string with $(n_1,n_2) =(1,0)$ and $(n_1,n_2)=(0,1)$ for $(q_{1},q_{2})=(4,1)$.

\section{Conclusions and Discussions}
\label{sec:conclusion}

In this paper, we studied the formation and the evolution of the string network in the Abelian Higgs model with two complex scalar fields. As a special feature of the model, the model possesses a global $U(1)$ symmetry in addition to the $U(1)$ gauge symmetry. Both symmetries are spontaneously broken by the VEVs of the two complex scalar fields. 

In this model, the dynamics of the string network is quite rich 
compared with that in the ordinary Abelian Higgs model with a single complex scalar field. 
In particular, we found the existence of a new type of string solutions,
the uncompensated strings, in addition to the ANO local string.
The isolated uncompensated string has a logarithmic divergent energy per unit length as in the case of the global strings, although it is accompanied by 
non-trivial gauge field configuration.

We also performed the classical lattice simulations in the $2+1$ dimensional spacetime, which confirmed 
the formation of the uncompensated strings at the phase transition.
We also found that most of the 
uncompensated strings evolve into the compensated strings
at later time when 
the gauge charge of the scalar field with 
the smaller VEV is larger than that of the scalar field with a larger VEV.
Such a behavior can be understood by 
long range forces between the uncompensated string and the global string.

Finally, we comment on the implications of the results in the present paper
to the axion models where $U(1)_{global}$
is identified with the Peccei-Quinn symmetry.
As we have mentioned in subsection\,\ref{sec:stringevolution},
we found that all the strings in our simulation for $(q_{1},q_{2})=(1,4)$ have the effective winding number either of $0$ or $\pm 1$ in the 
direction of the gauge-invariant Goldstone boson 
at late time.
When we apply this model to the axion model, 
at most one domain wall is attached to the cosmic string
below the QCD scale~\cite{Kawasaki:2013ae}).
Such string-wall network does not cause the infamous domain wall problem.
Thus, the axion model with $q_{1}=1$ and $q_{2}\ge 4$ and $\eta_1\gg \eta_2$ might be free from 
the domain wall problem even when
the $U(1)_{local}$ and $U(1)_{global}$
takes place after inflation.
The result of the present paper is, however, based on the simulation in the  $2+1$ dimensional spacetime, and hence, not conclusive.
We will perform the classical lattice simulation in the $3+1$ dimensional spacetime in a separate paper.

\vspace{-.4cm}  
\begin{acknowledgments}
\vspace{-.3cm}
MI thanks T.~Sekiguchi for useful discussion of the 
string formation.
This work is supported in part by JSPS KAKENHI Grant Nos. 15H05889, 16H03991, 17H02878, and 18H05542 (M.I.); 
World Premier International Research Center Initiative (WPI Initiative), MEXT, Japan (M.I.). 
\end{acknowledgments}

\appendix

\section{Domain Wall Problems in QCD Axion Models}
\label{sec:QCDaxion}
Here, we briefly summarize the axion domain wall problem
when we apply the accidental global symmetry,
$U(1)_{global}$, to the Peccei-Quinn symmetry~\cite{Peccei:1977hh,Peccei:1977ur,Weinberg:1977ma,Wilczek:1977pj} (see also \cite{Fukuda:2017ylt,Fukuda:2018oco,Ibe:2018hir}).
The Peccei-Quinn mechanism is one of the most successful solutions of the Strong CP problem,
where the effective $\theta$-angle of QCD is canceled by the VEV of the pseudo-Nambu-Goldstone boson, 
the axion (\cite{Kawasaki:2013ae} for review). 

To see how the axion appears in the present setup, 
let us consider the KSVZ axion model~\cite{Kim:1979if,Shifman:1979if} and introduce $N_1$ and $N_2$ of the vector-like 
multiplets of the fundamental representation 
of the $SU(3)_c$ gauge group of QCD which couple
to $\phi_1$ and $\phi_2^*$ through,
\begin{align}
   {\cal L}= \phi_1 \bar{Q}_1 {Q}_1  + \phi_2^*\bar{Q}_2 {Q}_2 + h.c.
\end{align}
Here, $(Q_i, \bar{Q}_i)$
collectively represents the $N_i$ vector-like 
multiplets ($i=1,2$).
We suppress the Yukawa coupling constants for notational simplicity.

Below the $U(1)_{local}$  and $U(1)_{global}$ breaking scales, the vector-multiplets become massive.%
\footnote{We assume that $\eta_{1,2}$ are much higher than the QCD scale.}
Then, by integrating out the vector-like multiplets, the couplings of the Goldstone bosons in Eq.\,\eqref{eq:decomp} to QCD
is given by,
\begin{align}
{\cal L} = \frac{g_s^2}{32\pi^2}
\left[\left( \frac{N_1 q_2 f_2^2 + N_2 q_1 f_1^2}{q_1^2 f_1^2+q_2^2 f_2^2}\right)
\frac{a}{F_a}
+ 
\left(N_1q_1 - N_2 q_2\right)\frac{b}{\sqrt{q_1^2 f_1^2+q_2^2 f_2^2}}
\right]G\tilde{G}\ .
\end{align}
Here,  $g_s$ is the QCD coupling constant, $G$ and $\tilde G$ are the gauge field strength of QCD and its dual.
We suppress the color and the Lorentz indices.
To satisfy the anomaly free condition of the $U(1)_{local}$ gauge symmetry, we require,
\begin{eqnarray}
N_1q_1 - N_2 q_2 = 0 \ ,
\end{eqnarray}
and hence, the effective Goldstone-QCD interactions 
are reduced to 
\begin{align}
{\cal L} = \frac{g_s^2}{32\pi^2}
\left[\frac{N_m a}{F_a}
\right]G\tilde{G}\ .
\label{eq:anomalous}
\end{align}
Here, we have introduced a parameter $N_m$ by $N_m  = N_1/q_2 = N_2/q_1$ ($N_m \in \mathbb{Z}^{>0}$).
Therefore, the gauge-invariant Goldstone can be identified with the QCD axion.

Now, let us discuss how the axion evolves in the cosmological history.
Let us assume that the $U(1)$ symmetry breaking takes place after inflation.
Then, the axion locally settles down to a point at the bottom of the scalar potential in Eq.\,\eqref{eq:potential}
after the phase transition.
The axion, however, winds around the cosmic string  with an effective winding number,
\begin{eqnarray}
n_{\rm eff} = n_1 q_2 - n_2 q_1 \ ,
\end{eqnarray}
in the domain of the axion $[0, 2\pi F_a)$ as given in Eq.\,\eqref{eq:domainA}.
Below the QCD scale, the axion coupling to QCD in Eq.\,\eqref{eq:anomalous} results in 
a periodic axion potential with a period, $F_a/N_m$.
Therefore, the axion potential gives rise the energy contrasts around the cosmic strings
and the cosmic strings are attached by $N_{\rm dw} = N_m \times |n_{\rm eff}|$ domain walls.
 For $|N_{\rm dw}| > 1$, the string-wall network is stable and immediately dominates the energy density of the universe, 
which conflicts with the Standard Cosmology..%
\footnote{For $N_m > 1$, the stability of the string-wall network is guaranteed by 
the discrete $\mathbb{Z}_{N_m}$ symmetry of the axion potential.
For $N_m = 1$, on the other hand, the string-wall network with $|n_{\rm dw}|>1$ can decay via quantum processes.
However, the rate is highly suppressed, and the lifetime of the string-wall network is much longer than 
the age of the universe.
}  

Around the compensated strings, i.e. $n_{\rm eff}=0$, on the other hand, no walls are formed, and hence, it does not 
cause cosmological problems.
Besides, the string-wall network with $N_{\rm dw} = 1$ also does not cause cosmological problems, 
since the network is not stable and decays immediately after the QCD phase transition~\cite{Vilenkin:1982ks,Hiramatsu:2012gg}.
The latter possibility requires $N_m = 1$ and $|n_{\rm eff}| = 1$.
Therefore, the domain wall problems in the axion model can be avoided if the cosmic strings 
evolves into either $R_{\rm c} \to 1$ or $R_{\rm dw} \to 0$ at a later time.%
\footnote{The domain wall problem can be also avoided when the Peccei-Quinn
symmetry breaking takes place before inflation.
In this case, the Hubble parameter during inflation is severely constrained 
by the isocurvature fluctuation of the axion dark matter.}

\bibliography{papers}

\end{document}